\documentclass[iop]{emulateapj}
\usepackage{txfonts}

\shorttitle{ChaMPlane bright X-ray sources}
\shortauthors{van den Berg et al.}

\begin{document}

\title{The {\em ChaMPlane} bright X-ray sources --- Galactic longitudes $\lowercase{l}=2^{\circ} - 358^{\circ}$} 

\author{Maureen van den Berg\altaffilmark{1,2}, Kyle
  Penner\altaffilmark{3}, JaeSub Hong\altaffilmark{2}, Jonathan
  E.~Grindlay\altaffilmark{2}, Ping Zhao\altaffilmark{2}, Silas
  Laycock\altaffilmark{4}, and Mathieu Servillat\altaffilmark{2}}
\affil{\altaffilmark{1}Astronomical Institute, Utrecht University,
  Princetonplein 5, Utrecht, 3508 TA, The Netherlands}
\affil{\altaffilmark{2}Harvard-Smithsonian Center for Astrophysics, 60
  Garden Street, Cambridge, MA 02138, USA; maureen@head.cfa.harvard.edu}
\affil{\altaffilmark{3}Department of Astronomy, University of Arizona,
  933 N. Cherry Ave., Tucson, AZ 85721, USA}
\affil{\altaffilmark{4}University of Massachusetts, Lowell, USA}


\begin{abstract}
The {\em Chandra} Multiwavelength Plane (ChaMPlane) Survey aims to
constrain the Galactic population of mainly accretion-powered, but
also coronal, low-luminosity X-ray sources ($L_{X} \lesssim 10^{33}$
erg s$^{-1}$). To investigate the X-ray source content in the plane at
fluxes $F_{\rm X} \gtrsim 3 \times 10^{-14}$ erg s$^{-1}$ cm$^{-2}$,
we study 21 of the brightest ChaMPlane sources, viz.~those with $>$250
net counts (0.3--8 keV). By excluding the heavily obscured central
part of the plane, our optical/near-infrared follow-up puts useful
constraints on their nature. We have discovered two likely accreting
white-dwarf binaries. CXOPS\,J$154305.5$$-$$522709$ (CBS 7) is a
cataclysmic variable showing periodic X-ray flux modulations on 1.2 hr
and 2.4 hr; given its hard spectrum the system is likely magnetic. We
identify CXOPS\,J$175900.8$$-$$334548$ (CBS 17) with a late-type
giant; if the X-rays are indeed accretion-powered, it belongs to the
small but growing class of symbiotic binaries lacking strong optical
nebular emission lines. CXOPS\,J$171340.5$$-$$395213$ (CBS 14) is an
X-ray transient that brightened $\gtrsim$100 times. We tentatively
classify it as a very late-type ($>$M7) dwarf, of which few have been
detected in X-rays. The remaining sources are (candidate) active
galaxies, normal stars and active binaries, and a plausible young
T\,Tauri star. The derived cumulative number density versus flux
($\log~N - \log~S$) relation for the Galactic sources appears flatter
than expected for an isotropic distribution, indicating that we are
seeing a non-local sample of mostly coronal sources. Our findings
define source templates that we can use, in part, to classify the
$>$10$^4$ fainter sources in ChaMPlane.
\end{abstract}

\section{Introduction}

The Galactic population of low-luminosity X-ray sources ($L_X\lesssim
10^{33}$ erg s$^{-1}$) includes normal and active single stars and
binaries, pre-main sequence stars, milli-second pulsars, and close
binaries containing accreting compact objects (white dwarfs in
cataclysmic variables or CVs; neutron stars and black holes in
quiescent high- and low-mass X-ray binaries, qHMXBs and qLMXBs). Due
to sensitivity constraints of previous surveys, little is known about
the distribution of these sources on Galactic scales, or about rare
sources for which one needs to sample a large volume to study them as
a class. With {\em XMM-Newton} and especially {\em Chandra}, which
combine sensitivity with excellent spatial resolution, existing
imaging Galactic X-ray surveys \citep[e.g.][]{hertgrin84} can be
extended with 2--3 orders of magnitude higher sensitivity down to
$10^{-15}$ erg s$^{-1}$ cm$^{-2}$ {\em and} with precise searches for
counterparts at other wavelengths. This is what drives several ongoing
campaigns, like the {\em XMM-Newton} Galactic Plane Survey
\citep[XGPS;][]{handea04} and our own {\em Chandra} Multi-wavelength
Plane survey \citep[ChaMPlane;][]{grinhongea05}. Such surveys are
important for understanding the X-ray emission from galaxies. What was
previously observed as a diffuse ``band'' of X-ray emission along the
plane (the Galactic Ridge X-ray Emission or GRXE) has, for a
significant part, been resolved into discrete sources \citep[about
  50\% at 3 keV above $10^{-16}$ erg s$^{-1}$ cm$^{-2}$ (0.5--7
  keV);][]{revnea09}. As the GRXE and the diffuse X-rays from
non-starbust galaxies have similar properties \citep{revnea08}, a
substantial point-source component is also implied for the latter. But
only in our own Galaxy can we study individual sources, as no
observatory in the foreseeable future is able to resolve the faint
X-ray emission from other galaxies.

\begin{table*}
\caption{The sample of bright sources \label{x-sample-table}}
\begin{center}
\begin{tabular}{clccrcccrcccc}
\hline
\hline
\multicolumn{1}{c}{(1)} & \multicolumn{1}{c}{(2)} & (3) & \multicolumn{1}{c}{(4)} & \multicolumn{1}{c}{(5)} & (6) & (7) & (8) & \multicolumn{1}{c}{(9)} & \multicolumn{1}{c}{(10)} & (11) & (12) & (13) \\
\multicolumn{1}{c}{CBS} & \multicolumn{1}{c}{CXOPS\,J} & ObsID- & \multicolumn{1}{c}{Date Obs} & T$_{\rm exp}$ & $N_{\rm H22,Gal}$ & $r_{95}$ & $\theta$ & \multicolumn{1}{c}{Net counts} & $F_{\rm X,lim}$ & Opt & Other & 2XMM\,J\\
 & & aimpoint & & \multicolumn{1}{c}{ks} & cm$^{-2}$ & $\arcsec$ & $\arcmin$ & &  & ID? & ObsID? & \\
\hline
 1 & $061759.2+222738$ & 04675-S & 04-04-12 & 58.3 & 0.22  & 0.31 & 1.24 & 318$\pm$19  &  5.3 & + & $-$ & 061759.1+222738 \\
 2 & $102801.4-435107$ & 03569-S & 03-05-23 & 26.5 & 0.07  & 0.40 & 4.06 & 258$\pm$17  &  7.7 & + & + & 102801.4$-$435106  \\
 3 & $104814.2-583051$ & 03842-S & 03-10-08 & 35.5 & 0.76  & 0.31 & 1.03 & 520$\pm$24  & 16.0 & + & $-$ & 104814.0$-$583051 \\
 4 & $111108.1-603657$ & 02782-S & 02-04-08 & 48.8 & 1.08  & 0.52 & 6.34 & 331$\pm$20  & 12.4 & + & $-$ & $-$ \\
 5 & $111149.5-604158$ & 02782-S & 02-04-08 & 48.8 & 1.15  & 0.32 & 2.41 & 899$\pm$31  & 12.4 & + & $-$ & 111149.6$-$604157 \\
 6 & $154204.8-522400$ & 00090-I & 00-04-08 & 23.6 & 1.09  & 0.36 & 3.24 & 351$\pm$20  & 31.9 & + & $-$ & 154205.0$-$522400 \\
 7 & $154305.5-522709$ & 00090-I & 00-04-08 & 23.6 & 1.26  & 0.53 & 7.41 & 669$\pm$27  & 31.9 & + & $-$ & 154305.5$-$522709 \\
 8 & $155052.4-562608$ & 05190-S & 03-10-23 & 47.7 & 0.58  & 0.33 & 2.77 & 491$\pm$23  & 10.5 & + & + & 155052.8$-$562609 \\
 9 & $155226.8-561704$ & 01965-S & 01-08-18 & 55.6 & 0.60  & 0.37 & 3.66 & 360$\pm$20  &  7.8 & + & $-$ & 155227.1$-$561702 \\
10 & $170000.9-265905$ & 01861-I & 01-07-04 & 32.3 & 0.14  & 0.46 & 7.52 & 2055$\pm$46\tablenotemark{a} & 11.6 & + & $-$ & $-$ \\
11 & $170905.5-443140$ & 04608-I & 04-02-01 & 97.2 & 0.91  & 0.63 & 7.55 & 299$\pm$19  &  6.7 & + & $-$ & 170905.5$-$443138 \\
12 & $170928.2-442916$ & 04608-I & 04-02-01 & 97.2 & 0.89  & 0.34 & 2.92 & 395$\pm$21  &  6.7 & + & + & 170928.1$-$442915 \\
13 & $170938.2-442255$ & 04608-I & 04-02-01 & 97.2 & 0.91  & 0.46 & 5.72 & 362$\pm$20  &  6.7 & $-$ & $-$ & 170938.1$-$442253 \\
14 & $171340.5-395213$ & 05559-I & 05-04-19 &  9.5 & 1.86  & 0.32 & 3.31 & 1740$\pm$43\tablenotemark{a} & 117 & + & $-$ & $-$ \\
15 & $171440.6-400234$ & 00737-I & 00-07-25 & 38.2 & 1.07  & 0.68 & 8.29 & 356$\pm$21     & 23.3 & $-$ & $-$ & 171440.7$-$400232 \\
16 & $171537.1-395559$ & 00737-I & 00-07-25 & 38.2 & 1.23  & 0.40 & 4.42 & 327$\pm$19     & 23.3 & $-$ & $-$ & 171537.0$-$395559 \\
17 & $175900.8-334548$ & 04586-S & 04-06-25 & 44.1 & 0.39  & 0.38 & 4.25 & 546$\pm$24   & 10.3 & + & $-$  & 175900.7$-$334547  \\
18 & $184355.1-035830$ & 02298-I & 01-05-20 & 96.6 & 5.82  & 0.52 & 7.03 & 520$\pm$24   & 20.5 & $-$ & + & $-$ \\
19 & $184421.1-035706$ & 949-I, 1523-I\tablenotemark{b} & 00-02-24 & 94.8 & 5.70 & 0.89 & 9.35 & 256$\pm$18  & 23.6 & + & $-$ & $-$  \\
20 & $222833.4+611105$ & 1948-I, 2787-I\tablenotemark{b} & 01-02-14 & 106.2 & 0.40 & 0.47 & 5.49 & 284$\pm$18 & 4.8 & + & $-$ & $-$ \\
21 & $235841.7+623437$ & 02810-I & 02-09-14 & 48.8 & 0.28  & 0.56 & 8.50 & 1168$\pm$35 & 9.1 & + & $-$ & $-$ \\
\hline
\end{tabular}

\tablenotetext{1}{Not corrected for pileup.}  \tablenotetext{2}{The
  source is detected in the stack of two observations. Exposure time,
  net counts, $r_{95}$, offset angle, and flux limit refer to the
  properties derived from the stacked data.}

\tablecomments{ Columns: 1) source number; 2) source name; 3)
  observation identification number and aimpoint; 4) date of start
  observation in format 20YY-MM-DD; 5) exposure time (GTI); 6)
  integrated Galactic column density in the direction of the source
  from \cite{drimea03} in units of 10$^{22}$ cm$^{-2}$; 7) radius of
  the 95\% error circle; 8) angular offset from the aimpoint; 9) net
  counts (0.3--8.0 keV); 10) flux limit in units of $10^{-14}$ erg
  s$^{-1}$ cm$^{-2}$ (0.3--8 keV, unabsorbed) for a source at the
  aimpoint that corresponds to a minimum of 250 net counts, for a
  power-law spectrum with $\Gamma=1.7$ and $N_{\rm H}=N_{\rm H,Gal}$
  (see column 6); 11) flag for the detection of a candidate optical
  counterpart; 12) flag for the detection in other ACIS pointings in
  our database. The associated ObsIDs are for CBS 2: 835; CBS 8: 1845,
  1846, 3672, 3807; CBS 12: 757; CBS 18: 949, 1523; 13) potential 2XMM
  counterpart.}
\end{center}
\end{table*}

The main goal of ChaMPlane is to constrain the Galactic distribution
of faint accretion-powered sources (mainly CVs); the secondary goal is
to do a deep survey of stellar coronal sources. We systematically
process {\em Chandra} Advanced CCD Imaging Spectrometer
\citep[ACIS;][]{garmea03} pointings in the plane to analyze
serendipitous detections, and do optical and near-infrared (nIR)
follow-up for source classification. Whereas the XGPS studies a
specific, approximately 3-deg$^2$ region of the plane between Galactic
longitudes $l=19^{\circ}$ and 22$^{\circ}$, ChaMPlane uses archival
data without any restrictions on longitude. By now, our coverage is
about 7 deg$^2$ and encorporates our dedicated surveys of
low-extinction bulge regions \citep[``Windows Survey'';
  e.g.][]{vdbergea09,hongvandea09} and of a latitudinal strip around
the Galactic center (``Bulge Latitude Survey''; Grindlay et al.~2012,
in preparation). Our focus on the bulge is driven by trying to uncover
the nature of the thousands of hard inner-bulge sources that are
believed to be accreting compact objects---the largest such population
in the Galaxy \citep[see e.g.][]{munoea09}.

Most of our sources are detected with only a few tens of counts or
less. Here we want to highlight the brightest ChaMPlane sources, for
which we can study the X-ray spectra and light curves in more
detail. Not only does this sample include objects that are worth
further investigation on their own account, but it also allows us to
study the typical content and flux distribution of sources covered by
the high-flux end of our survey ($F_{\rm X} \gtrsim 3 \times 10^{-14}$
erg s$^{-1}$ cm$^{-2}$), and define characteristic source types to be
used as templates for the many faint sources in our database.  Despite
our focus on the bulge, for a survey like ChaMPlane this region also
has its disadvantages. The high stellar density complicates source
identification in the optical/nIR. Moreover, the severe extinction
along the line of sight restricts a priori the detection of
intrinsically faint (in the optical/nIR) objects like CVs to a few
kpc, as illustrated by our CV discoveries in a few central-bulge
fields \citep{koenea08}. Therefore, we defer the study of our
brightest sources towards the central part of the Galaxy to a future
paper.  Here we exclude the central $\pm$2$^{\circ}$ around the
Galactic center.

In \S\ref{data-analysis} we describe the sample selection and data
analysis. We present the sources by class in \S\ref{results}. In
\S\ref{discuss} we place a few individual sources in a broader
context, and consider the sample as a whole.  Preliminary results were
reported in \cite{pennea08}.

\section{Sample selection and data analysis}\label{data-analysis}

At this writing, the ChaMPlane X-ray source catalog contains
$\sim$15\,000 sources from archival ACIS imaging data at Galactic
latitudes $|\,b\,|<12^{\circ}$ that meet the primary criteria
established in \cite{grinhongea05}: preferably ACIS-I exposures (for
the larger field of view) with exposure times $\gtrsim 20$ ks and
without bright or extended targets that limit the sensitivity to
serendipitous detections. Our preference for fields with a minimal
column density $N_{\rm H}$ is almost impossible to realize in the
bulge. In \S\ref{sample-criteria} we summarize the additional criteria
adopted to select a bright-source sample, and we explain the X-ray and
optical/nIR data analysis in \S\ref{x-sample} and
\S\ref{optical-sample}.  General classification guidelines are
outlined in \S\ref{source_class}. Cross-correlation of the sample with
other X-ray catalogs is described in \S\ref{sec_xcat}.

\subsection{Sample selection criteria}\label{sample-criteria}

Three selection rules define the current sample. First, the number of
net counts between 0.3 and 8 keV (our $B_X$ band) has to exceed 250 so
that a meaningful fit to the X-ray spectrum can be made. Second, to
keep positional errors small, a source cannot lie too far from the
aimpoint.  We only select detections on CCDs I0--I3 and CCD S3 for
observations that use the ACIS-I and ACIS-S aimpoint,
respectively\footnote{See Fig.~6.1 in the {\em Chandra Proposers'
    Observatory Guide} (POG) at
  http://cxc.harvard.edu/proposer/POG/html for the layout of ACIS
  detector array.}. For a source with 250 net counts or more, this
results in 95\% confidence radii that are $\lesssim$ 1\farcs0 for
aimpoint offset angles up to 10\arcmin\,(which covers the entire S3
chip and most, i.e.~$\sim$92\%, of the ACIS-I array) and up to
1\farcs6 out to the extreme corners of ACIS-I. Positional errors are
estimated using the empirical relation between the 95\% confidence
radius, net counts, and offset angle presented in \cite{hongvandea05},
which is based on extensive MARX simulations.  Finally, we exclude the
region within 2$^{\circ}$ of the Galactic center (GC) to avoid the
most crowded and obscured part of the plane. A total of 63
observations covering 3 deg$^2$ satisfy this
criterion. \cite{zhaogrindea05} includes a list of our fields that
cover these observations; a few more were added recently, but most of
these lie within 2$^{\circ}$ of the GC and would not have been
considered for the present study. The only field that is included here
but is not listed in \cite{zhaogrindea05} is located at
($l$,$b$)=(148.19$^{\circ}$,+0.81$^{\circ}$) and was covered by an
ACIS-S observation.
 
\begin{figure}
\centering
\includegraphics[width=8cm]{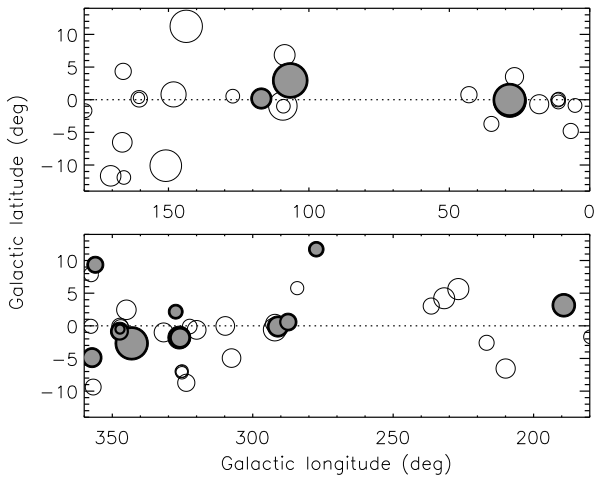}
\caption{Distribution of the fields from which we selected the bright
  sources. Filled circles are the fields that include targets from
  Table~\ref{x-sample-table}. The symbol size scales with the exposure
  time, which has a maximum GTI value of 106.2 ks. \label{fig-dist}}
\end{figure}

Our final sample contains 21 sources from 14 single observations and 2
stacked fields that each consist of 2 overlapping observations added
together to increase sensitivity (\cite{hongvandea05} give details of
our stacking procedure). Table \ref{x-sample-table} lists the sources
in order of increasing right ascension, and summarizes the basic
properties. Fig.~\ref{fig-dist} shows their distribution in the
plane. For convenience we use the acronym CBS (ChaMPlane Bright
Source) to designate the sources in the text instead of their CXOPS
name (column 2 of Table \ref{x-sample-table}).

\begin{table*}
\begin{center}
\caption{Parameters of the best model fits to {\em Chandra} spectra \label{spectral-param}}
\begin{tabular}{ccccccc}
\hline
\hline
\multicolumn{7}{c}{Power-law model}\\
\cline{1-7}
CBS & $\Gamma$                     & $N_{\rm H,X}$       & $Z/Z_{\odot} $        & $F_{\rm 0.3-8,u}$ & $F_{\rm 2-8,u}$ & $\chi^{2}_{\nu}$/d.o.f \\
    &                              & $10^{22}$ cm$^{-2}$ &                     & 10$^{-14}$ erg s$^{-1}$ cm$^{-2}$ & 10$^{-14}$ erg s$^{-1}$ cm$^{-2}$ &      \\                       
\hline
1   & 1.7$^{+0.4}_{-0.3}$ & 0.7$\pm$0.3    & N/A   & 9.3$\pm$0.6 & 5.0$\pm$0.3                   &  1.60/12                \\
2   & 1.9$\pm$0.3       & 0.15$\pm$0.09  &N/A     & 8.5$\pm$0.6 & 3.8$\pm$0.3                     &  1.82/8             \\
7   & 1.0$\pm$0.1       & 0.2$\pm$0.1    &N/A     & 51$\pm$2    & 40$\pm$2                     &  0.94/28               \\
9   & 0.7$\pm$0.3       & 2.2$\pm$0.8    &N/A     & 21$\pm$1    & 18$\pm$1                     &  0.93/14               \\
13  & 2.0$\pm$0.5       & 6.2$\pm$1.6    &N/A     & 36$\pm$2    & 16$\pm$1                     &  1.26/15         \\
15  & 1.9$\pm$0.4       & 1.5$\pm$0.4    &N/A     & 41$\pm$2    & 19$\pm$1                     &  0.74/14               \\
16  & 0.9$\pm$0.3       & 0.9$\pm$0.6    &N/A     & 23$\pm$1    & 18$\pm$1                 &  1.29/11                 \\
17  & 0.8$\pm$0.2       & 1.0$\pm$0.3    &N/A     & 52$\pm$2    & 43$\pm$2                 &  0.91/23                \\
18  & $-$0.26$\pm$0.4   & 2.1$^{+1.5}_{-1.3}$ &N/A & 23$\pm$1    & 22$\pm$1                 & 1.47/22                  \\
\hline
\multicolumn{7}{c}{Thermal models} \\
\cline{1-7}
CBS & $kT$                         & $N_{\rm H,X}$       & $Z/Z_{\odot} $        & $F_{\rm 0.3-8,u}$ & $F_{\rm 2-8,u}$ & $\chi^{2}_{\nu}$/d.o.f \\
    & keV                          & $10^{22}$ cm$^{-2}$ &                     & 10$^{-14}$ erg s$^{-1}$ cm$^{-2}$ & 10$^{-14}$ erg s$^{-1}$ cm$^{-2}$ &      \\                       
\hline                                                                                                                                     
3   & 0.60$\pm$0.07                & 0.11$\pm$0.04     & 0.04$\pm$0.01        & 10.6$\pm$0.5  & 0.39$\pm$0.02   &                           0.89/20 \\              
4   & 4$\pm$1             &  0.22$\pm$0.06   & $\equiv$1    & 7.1$\pm$0.4                  & 3.8$\pm$0.2   & 1.11/12         \\
5   & 0.18$\pm$0.02, 0.58$\pm$0.07 & 0.46$\pm$0.07       & $\equiv$1   & 80$\pm$3         & 0.24$\pm$0.01   & 1.10/28          \\ 
6   & 0.7$\pm$0.1                  & 0.13$\pm$0.11     & 0.10$^{+0.07}_{-0.04}$   & 17$\pm$1             & 0.70$\pm$0.04   & 1.05/11 \\           
8   & 5.5$^{+2.3}_{-1.3}$             & 0.25$\pm$0.08    & N/A & 11.1$\pm$0.5                 & 6.0$\pm$0.3   & 1.02/18 \\           
10  & 0.31$\pm$0.05, 1.08$\pm$0.04 & $<$0.05           & 0.18$\pm$0.03            & 67$\pm$1\tablenotemark{a} & 6.7$\pm$0.2\tablenotemark{a}   &     1.47/57             \\

11  & 0.9$\pm$0.1                  &  $<$0.07  & 0.11$\pm$0.06            & 3.0$\pm$0.2                  & 0.25$\pm$0.02   &  0.82/10             \\
12  & 0.8$\pm$0.1                  &  $<$0.18  & 0.07$\pm$0.05            & 3.5$\pm$0.2                  & 0.22$\pm$0.01   &  0.74/12             \\

14  & 0.9$\pm$0.1, 2.5$\pm$0.3     & $<$0.02           & 0.5$^{+0.3}_{-0.2}$      & 340$\pm$8\tablenotemark{a} &  112$\pm$3\tablenotemark{a}   & 1.01/28    \\       
19  & 4$^{+2}_{-1}$                  & 0.7$\pm$0.3       & $\equiv$1    & 5.7$\pm$0.4                  & 2.8$\pm$0.2    & 0.81/10 \\         
20  &  1.0$^{+0.1}_{-0.2}$             & $<$0.29           & 0.04$^{+0.06}_{-0.03}$            & 2.6$\pm$0.1                  & 0.31$\pm$0.02   &   0.94/6   \\           
21  & 0.79$\pm$0.03                 & $<$0.02          & 0.10$\pm$0.02            & 22.3$\pm$0.7                 & 1.50$\pm$0.05   &      1.42/41 \\            
\hline
\end{tabular}

\tablenotetext{1}{After correcting for pile up. The fluxes for CBS 14
  are very uncertain given the high pileup fraction, see text.}

\tablecomments{{\em Top:} Sources with spectra that are best fit with
  a power law. {\em Bottom:} Sources with spectra that are best fit
  with a thermal model, which is a 1$T$ or 2$T$ MeKaL plasma except
  for CBS 8 where thermal bremsstrahlung provides a better fit.  From
  left to right: source number; photon index $\Gamma$ or plasma
  temperatures $kT$; column density; metal abundance (only for MeKaL
  models); unabsorbed flux in the 0.3--8 and 2--8 keV bands; reduced
  $\chi^2$ and degrees of freedom. Maximum values for $N_{\rm H,X}$
  correspond to 1-$\sigma$ upper limits. Errors on the flux only take
  into account the statistical error in the net counts
  (Table~\ref{x-sample-table}).}
\end{center}
\end{table*}

\subsection{X-ray analysis}\label{x-sample}

\nocite{hongvandea05,hongvandea09} Hong et al.~(2005, 2009) explain
the ChaMPlane pipeline for processing the ACIS data.  We refer to
those papers for details on source detection, and the computation of
net source counts and 95\% confidence radii on source positions
($r_{95}$). We compute energy quantiles $E_{x}$ following the method
described in \cite{hongschlea04}, which allows us to study the
time-resolved spectral properties of our sources.

For each source we create a background-subtracted lightcurve with the
CIAO tool {\em dmextract}. Counts are extracted from a region that
includes 95\% of the energy at 1.5 keV; the background is estimated
from a source-free annular region centered on the source position. A
Kolmogorov-Smirnov (K-S) test on photon arrival times indicates that
three sources may be variable (probability for a constant count rate
$p_{KS} <1$\%).  Since the K-S test is only a rough guide to source
variability, we follow up with visual inspection of each light
curve. We thus identify CBS 7 as a possible fourth variable ($p_{KS} =
26$\%). The variability of these sources is further discussed in
\S\ref{results}.

We use \emph{psextract} to extract spectra from the same regions used
to measure the net source counts, and generate rmf and arf response
files for the source and background regions. For the purpose of
fitting, spectra are grouped with at least 20 counts per bin in order
to use the $\chi^2$ statistic.  Data that have accumulated in the
ChaMPlane archive have been processed with various CIAO versions and
calibration files. For consistency, we use the more recently
calibrated data from the {\em Chandra} archive (processed with
CIAO\,3.4/CALDB\,3.30) for the spectral analysis of all
sources. Spectra are fitted within {\tt Sherpa}\footnote{See
  http://cxc.harvard.edu/sherpa/.} version 4.3 to a non-thermal and
two thermal models: a power law (the \emph{powlaw1d} model), thermal
bremsstrahlung (\emph{xsbremss}), and a MeKaL model for an
optically-thin thermal plasma appropriate for stellar coronae
(\emph{xsmekal}). For the MeKaL model, we fitted one (1$T$) and
two-temperature (2$T$) plasmas, and treated the global metal abundance
$Z$ as a fit parameter or fixed it to the solar value. We only adopt
the results of the variable-$Z$ or 2$T$ model if fitting for the extra
parameters significantly improves the fit\footnote{F-test probability
  of the more complex model being correct $>$95\%.}; based on this
criterion a 2$T$ model turns out to be warranted for 3 sources. We
account for an absorbing column density $N_{\rm H}$ using the
\emph{xsphabs} model. For each source, the best-fit model is presented
in Table \ref{spectral-param}.  All errors quoted correspond to 68\%
(1$\sigma$) confidence intervals, computed with the {\em confidence}
method inside {\tt Sherpa}. For CBS 20, we report the results of
fitting the spectrum of ObsID 2787 only, which constitutes $\sim$95\%
of the total spectrum.

\begin{table*}
\caption{Optical and nIR properties of the candidate optical counterparts}
\begin{center}
\begin{tabular}{crrrrrrrrrrl}
\hline
\hline
(1) & \multicolumn{1}{c}{(2)} & \multicolumn{1}{c}{(3)} & \multicolumn{1}{c}{(4)} & \multicolumn{1}{c}{(5)} & \multicolumn{1}{c}{(6)} & \multicolumn{1}{c}{(7)} & \multicolumn{1}{c}{(8)} & \multicolumn{1}{c}{(9)} & \multicolumn{1}{c}{(10)} & \multicolumn{1}{c}{(11)} & (12) \\
CBS & \multicolumn{1}{c}{d{$_{X-O}$}} & \multicolumn{1}{c}{d{$_{X-O}$}} & \multicolumn{1}{c}{$N_{\rm ran}$} &  \multicolumn{1}{c}{$V$} &  \multicolumn{1}{c}{$R$} &  \multicolumn{1}{c}{$I$} &  \multicolumn{1}{c}{$H\alpha - R$} & \multicolumn{1}{c}{$J$} & \multicolumn{1}{c}{$H$} & \multicolumn{1}{c}{$K_s$} & Spectral type  \\
    & \multicolumn{1}{c}{\arcsec}   & \multicolumn{1}{c}{$\sigma$}  \\
\hline
1                  & 0.19 & 0.96 & 0.02  & 22.9(2)    & 21.9(1)  & 20.3(1)  & $-0.1(2)$   & \nodata  & \nodata  &\nodata   & \nodata                                         \\
2                  & 0.36 & 1.53 & 0.04  & 19.2(1)    & 18.9(1)  & 18.4(1)  & $0.0(1)$     & \nodata  & \nodata  &\nodata   & AGN ($z=1.78$)                  \\ 
3                  & 0.08 & 0.41 & 0.04  & 13.4(1)   & 12.5(1) & 11.7(1) & $0.0(1)$   & 11.36(2) & 10.86(3) & 10.69(3) & mid K; ${\rm EW_{H\alpha}} \approx -0.2\pm0.2$ \AA                                            \\
4                  & 0.25 & 0.89 & 0.09  & \nodata    & 17.3(1)  & \nodata  & $-0.1(1)$    & 15.26(6) & 14.49(6) & 14.29(7) & mid K; ${\rm EW_{H\alpha}} \approx -3.4\pm0.4$ \AA\tablenotemark{c}                                            \\
5                  & 0.16 & 0.88 & 0.001 &  8.09\tablenotemark{a} &  7.95\tablenotemark{a}    &  7.79\tablenotemark{a}    & \nodata        &  7.66(3) &  7.65(4) &  7.62(3) & O7.5III(n)((f))    \\
6                  & 0.42 & 1.97 & 0.10  & \nodata    & $\la 17$\tablenotemark{b} & \nodata & \nodata        & 10.73(3) & 10.40(3) & 10.35(3) & \nodata                                         \\
7                  & 0.48 & 1.94 & 0.17  & 21.5(1)    & 20.9(2)  & 20.3(1)  & $-0.5(2)$    & \nodata  & \nodata  &\nodata   & \nodata                                          \\
8                  & 0.18 & 0.86 & 0.06  & 14.5(1)   & 13.3(1) & 12.2(1) & $-0.1(1)$    & 10.43(2) &  9.57(2) &  9.32(2) & early/mid K, filled-in H$\alpha$?                                     \\
9                  & 0.31 & 1.43 & 0.08  & 22.5(1)    & 20.9(1)  & 19.4(1)  & 0.0(1)        & \nodata  & \nodata  &\nodata   & \nodata                                          \\
10                 & 0.73 & 2.81 & 0.16  & 12.5(1)   & \nodata  & \nodata  & \nodata        &  9.80(3) &  9.13(2) &  8.95(2) & \nodata                                          \\
11                 & 0.28 & 1.00 & 0.33  & \nodata    & 14.0(1) & \nodata  & $-0.2(1)$   & 12.23(3) & 11.56(4) & 11.44(4) & mid/late K; ${\rm EW_{H\alpha}} \approx -1.1\pm0.3$ \AA              \\
12                 & 0.17 & 0.99 & 0.13  & \nodata    & 12.9(1) & \nodata  & 0.1(1)       & 11.91(5) & 11.58(4) & 11.47(3) & early/mid K                                      \\
14                 & 1.38 & 2.88 & 0.50  & 21.6(1)    & 18.7(1)  & 16.6(1)  & 0.2(1)        & 13.40(3) & 12.56(3) & 12.10(3) & \nodata                                          \\
17                 & 0.14 & 0.62 & 0.19  & 15.4(1)   & 14.0(1) & \nodata  & $-0.1(1)$   & 11.01(3) &  9.99(3) &  9.72(3) & K7--M1 III; ${\rm EW_{H\alpha}} \approx -2.7\pm0.4$ \AA              \\
19                 & 1.34 & 2.88 & 0.23  & 18.0(1)    & 16.2(1)  & 14.9(1) & $-0.1(1)$     & 12.73(3) & 11.72(3) & 11.04(3) & young star; ${\rm EW_{H\alpha}} \approx -6.6\pm0.9$ \AA\\
20                 & 0.60 & 2.25 & 0.04  & 15.6(1)   & 14.8(1) & \nodata  & $-0.1(7)$   & 12.93(3) & 12.28(3) & 12.16(3) & early/mid K; filled-in H$\alpha$?                                     \\
21                 & 0.49 & 1.60 & 0.07  & 12.2(1)   & \nodata  & 11.3(1) & \nodata        & 10.68(2) & 10.23(3) & 10.14(2) & mid G, early K; filled-in H$\alpha$?                                     \\
\hline
\end{tabular}

\tablenotetext{1}{Optical photometry is taken from \cite{vazqfein90};
  we assume magnitude errors of 0.05. The spectral type is taken from
  \cite{walb73}, to which we refer for an explanation of the
  qualifiers ``(n)'' and ``((f))''.}  \tablenotetext{2}{This star is
  overexposed in our deep Mosaic images, and falls in a chip gap in
  the shallow images.}  \tablenotetext{3}{Could be residual H$\alpha$
  emission from the background.}

\tablecomments{ Columns: 1) ChaMPlane Bright Source number; 2) angular
  separation between X-ray source and candidate optical counterpart
  {\em after} correction for boresight; 3) same, but in units of the
  X-ray/optical match radius $\sigma$; 4) number of spurious matches
  based on the local projected source density down to the limiting
  magnitude of the Mosaic catalog; 5--8) optical photometry; 9--11)
  2MASS photometry; 12) spectral classification, and, where relevant,
  equivalent width (EW) of the H$\alpha$ emission line derived from
  optical spectra. Errors on the last significant digit are given in
  parentheses. \label{opt-sample-table}}

\end{center}
\end{table*}

CBS 10 and 14 are piled up by 6\% and 30\%, respectively. We corrected
the count rates accordingly, but the correction for CBS 14 is quite
uncertain\footnote{See the {\em Chandra ABC Guide to Pileup}. We
  assumed that the grade mitigation parameter $\alpha=0.8$. CBS 10 is
  only mildly piled up and the result is insensitive to the value of
  $\alpha$. For CBS 14, the pileup fraction is 67\% (25\%) for
  $\alpha=0.75$ (0.9); the quoted fluxes change by 70\% for this range
  of $\alpha$.}. To make sure that the spectral fits are not affected
by pileup, we excluded events from the central source regions with
radius $r$. By trying different values of $r$, we found that $r=1.5$
(CBS 10) and $r=2.0$ (CBS 14) pixels are adequate (i.e.~larger values
give the same fit results) leaving 90\% and 50\% of the detected
events available for spectral fitting.  Energy fluxes were computed by
applying the derived count rate-to-flux conversion factors to the
pileup-corrected count rates. For CBS 10 the 2$T$ MeKaL fit seems to
systematically underestimate the count rate in the highest energy bins
between 2 and 4 keV even after removing a large core region. A third,
hotter component may be needed to improve the fit.

Column 12 of Table~\ref{x-sample-table} indicates whether a source is
detected in a pointing included in our database other than the one
listed in column 3.  By definition these additional detections have
fewer counts. By comparing count rates and energy quantiles (0.3--8
keV) of each detection, we found that CBS 2, 8 and 12 are potentially
long-term variables. This is discussed further in \S~\ref{results}.

\subsection{Optical/nIR data and analysis}\label{optical-sample}

Each ChaMPlane field is imaged through the $VRI$H$\alpha$ filters
using the Mosaic cameras on the KPNO-4m and CTIO-4m
telescopes. Sequences of deep and shallow exposures provide coverage
from $R \approx 12$ to 24 with $S/N \gtrsim 5$.  Our values for
H$\alpha - R$ colors are defined as the offset from the median
H$\alpha - R$ color of all stars in a given field; negative values
mark an H$\alpha$ excess flux. We tie the absolute astrometry of the
Mosaic images to the International Celestial Reference System (ICRS)
using stars in the USNO-A2 \citep{moneea98} or 2MASS \citep{skruea06}
catalogs with rms residuals to the fit of typically $\lesssim
0\farcs1$. The astrometric solution of Chandra images as a whole can
be offset from the ICRS by up to 0\farcs7 (90\% uncertainty\footnote{
  http://cxc.harvard.edu/cal/ASPECT/celmon/}). Without correcting for
such a systematic offset, or {\em boresight}, positions of X-ray
sources could be shifted with respect to those of their true optical
counterparts with an amount that is much larger than $r_{95}$, which
would harm the optical identification.  After correcting for the
boresight, we cross-correlate the X-ray and optical source catalogs to
look for candidate optical counterparts. The adopted $3\sigma$ search
radius takes into account errors on the X-ray and optical positions as
well as the boresight error. \citet{zhaogrindea05} describe the
details of our optical imaging campaign, the image processing,
computing the boresight correction, and the matching procedure.

ChaMPlane's spectroscopic campaign is still ongoing. For many of the
candidate optical counterparts in Table~\ref{x-sample-table} we have
already obtained low-resolution (3--7 \AA) optical spectra during
various runs conducted between 2002 and 2010 with the FAST
spectrograph on the FLWO-1.5m telescope, the IMACS spectrograph on the
6.5m Magellan Baade telescope, and the Hydra spectrographs on the
WIYN-3.5m and CTIO-4m telescopes. The spectra are reduced and
extracted with a combination of standard IRAF software and dedicated
packages. We assign spectral types by comparison with spectral
standards observed with a similar resolution
(e.g.~\citealt{jacoea84}). As a check we run the {\tt SPTCLASS} code
by
J.~Hernandez\footnote{http://www.astro.lsa.umich.edu/~hernandj/SPTclass/sptclass.html},
which assigns spectral types by measuring the strength of certain
absorption features. Both methods give consistent results.

\begin{figure*}
\centering
\includegraphics[width=7.75cm]{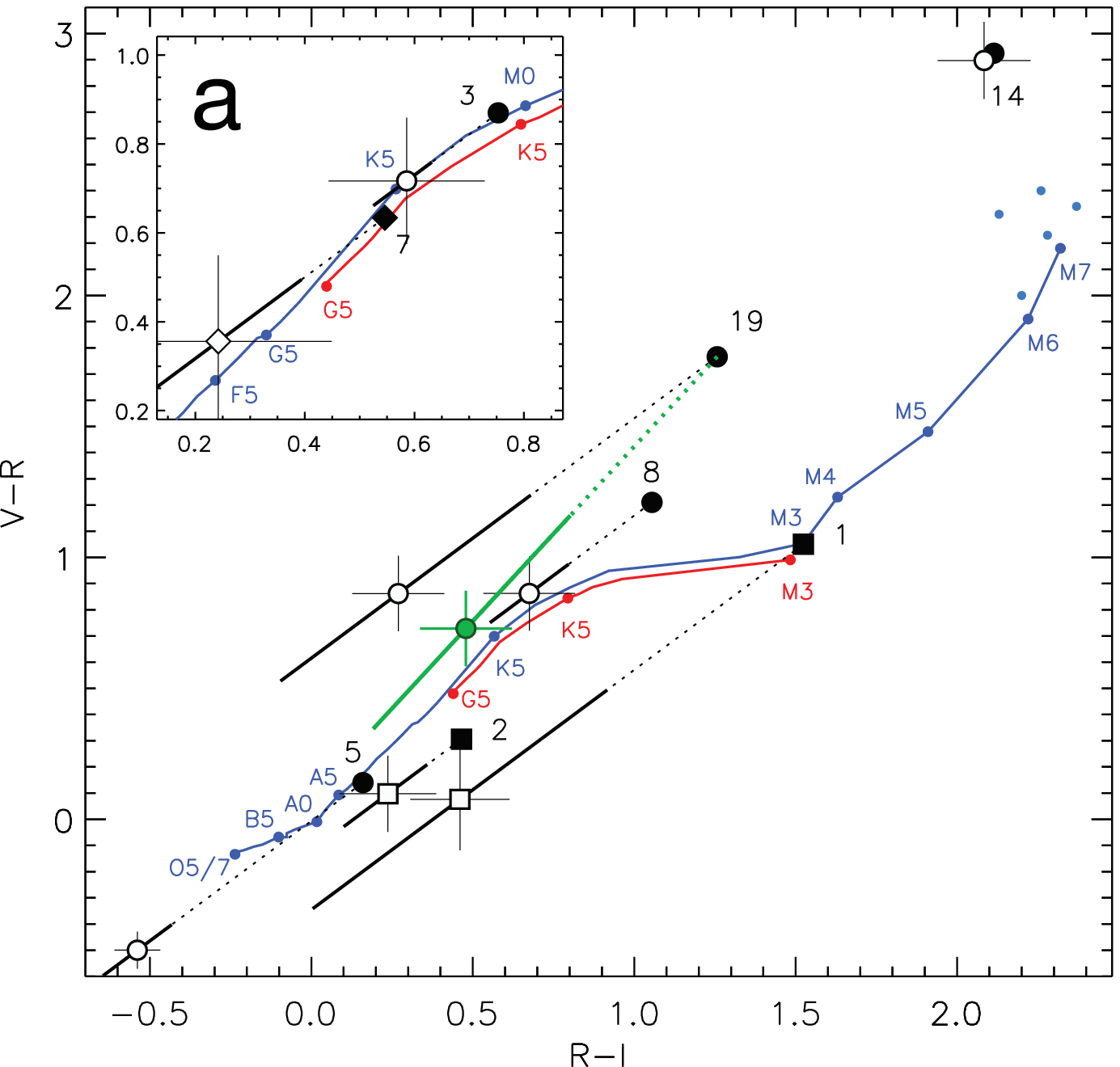}
\hspace{0.2cm}
\includegraphics[width=8.15cm]{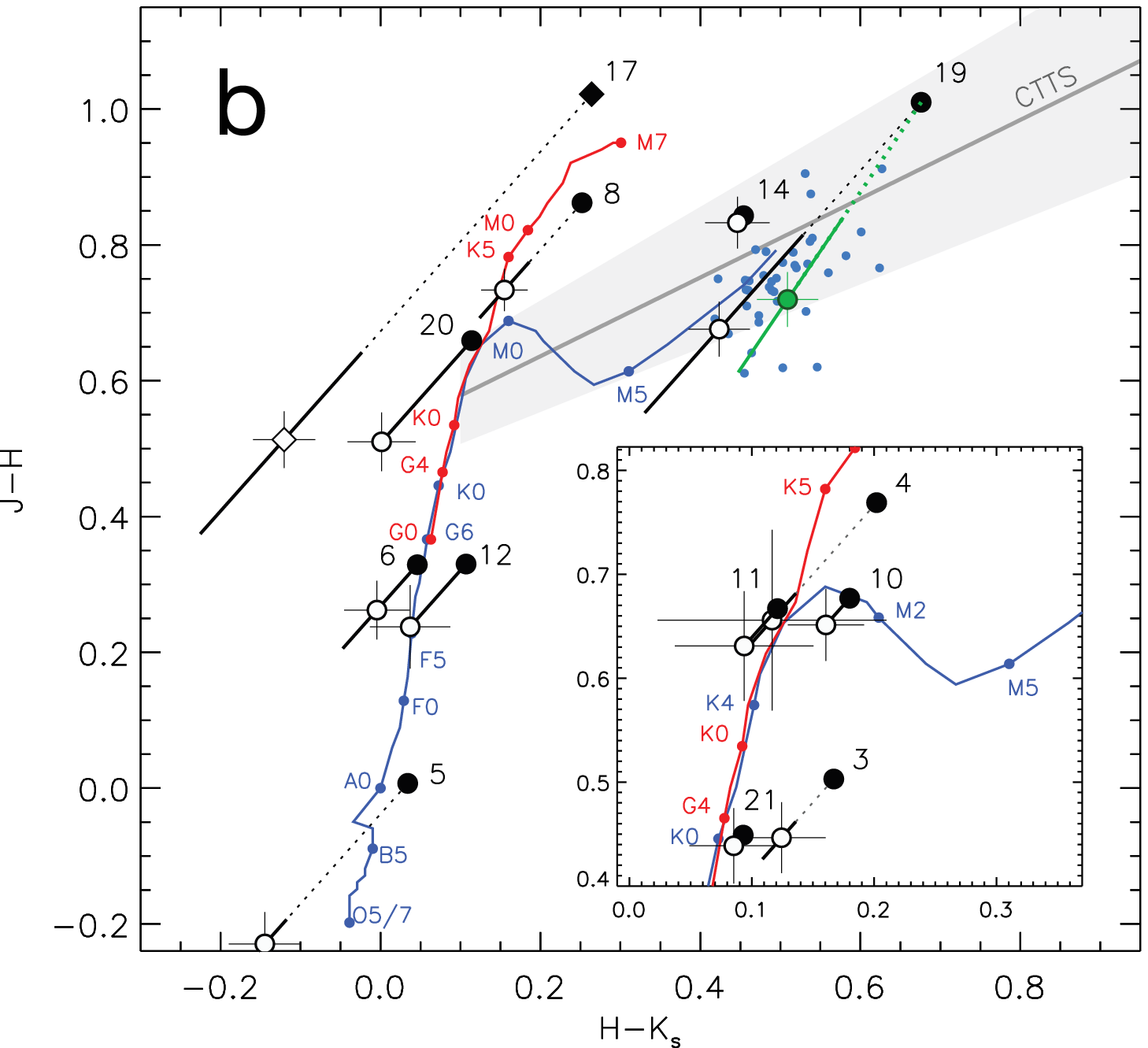}
\caption{ Optical ({\em left}) and nIR ({\em right})
    color-color diagrams showing the observed colors of the candidate
    counterparts ({\em filled black symbols}) connected with a dotted
    line to their dereddened colors ({\em open symbols}). Error bars
    are only plotted on the latter. The lengths of the reddening
    vectors are based on $N_{\rm H,X}$, with the solid thick part of
    the lines representing the 1-$\sigma$ errors. The slope of the
    vectors follows the extinction coefficients from
    \cite{cardea89}. As an example of extinction laws that are more
    appropriate for the bulge, we include ({\em in green}) a reddening
    vector based on \cite{popoea03} and \cite{sumi04} (for the
    optical), and \cite{nishea08} (for the nIR).  The gray area in the
    right panel marks the locus of classical T~Tauri stars (CTTS)
    \citep{meyeea97}.  Circles mark sources we classify as stars,
    squares are likely AGNs, and diamonds are likely accreting
    binaries. The counterpart for the candidate AGN CBS 9 is not
    plotted for clarity. The blue and red lines are the intrisic
    colors of main-sequence stars and giants, respectively, taken from
    \cite{john66}, \cite{bess91}, and \cite{bessbret88}. Filled blue
    circles show the colors of late-type M-dwarfs taken from Henry et
    al.~(2004; spectral types M\,8, M\,9, M\,9.5) and Cruz et
    al.~(2003; spectral types M\,9, L\,0, L\,1). The insets zoom in on
    crowded regions. {\em See the electronic edition of the Journal
      for a color version of this figure.}
\label{fig_ccd}}
\end{figure*}

In summary, of the 21 sources in our sample 17 have candidate optical
counterparts, and in each case only a single match is found.  Two of
these (CBS 5 and 6) have very bright optical matches that are
overexposed in the Mosaic images; for CBS 5, we use photometry from
the literature. We have optical spectra for 12 of the 17 candidate
counterparts. Table \ref{opt-sample-table} lists the properties of all
candidate counterparts. The optical/nIR color-color diagrams of
Fig.~\ref{fig_ccd} are used for constraining luminosity classes and
classifications when optical spectra are not available; see the
discussion of individual sources in \S~\ref{results}.

\nocite{henrea04,cruzea03}

Finding an optical match does not guarantee it is the true
counterpart, unless it shows other signs of activity (e.g.~excess
H$\alpha$ flux) in which case a physical association is very
likely. The random-match probability depends on several factors,
including the local projected star density and the separation between
the X-ray and optical source. To give an idea of the expected number
of chance coincidences, we give in column 4 of
Table~\ref{opt-sample-table} the number of stars expected in the
3$\sigma$ search area based on the optical source density (down to the
limiting magnitude of the Mosaic images) within 1\arcmin~of the
source.

To complement the optical photometry we looked up the $JHK_s$
magnitudes of the optical matches in the 2MASS catalog and included
them in Table \ref{opt-sample-table}. The quality flags indicate that
the $J$ and $H$ magnitudes for CBS 19 are potentially contaminated by
an image artifact. This star is also included in the UKIDSS Galactic
Planey Survey (GPS) catalog \citep{lucaea08}; comparison of the $J$
and $H$ magnitudes shows them to be very similar to the 2MASS values
with differences in the nIR colors of only $\sim$0.1 mag. None of the
other sources appear in the GPS catalog, which covers a limited
portion of the plane and is still in progress.

\subsection{Source classification}\label{source_class}

A useful diagnostic to discriminate between stellar coronal sources,
accreting binaries, and active galactic nuclei (AGN) is the ratio of
the intrinsic and unabsorbed (u) X-ray and optical flux: $\log (F_{\rm
  X}/F_{\rm O})_{\rm u} = \log F_{\rm X,u} + m/2.5 - {\rm ZP}.$ Here,
$m$ is the unabsorbed optical magnitude, and ZP is the logarithm of
the optical flux for a star with magnitude 0. Assuming a filter width
of 1000 \AA, the value of ZP is $-5.44$ for the $V$ band, and $-5.66$
for the $R$ band \citep{bessea}. Most coronal sources have $\log
(F_{\rm X}/F_{\rm O})_{\rm u} \lesssim -1$, while accretion-powered
sources typically have higher values \cite[e.g.][]{stocea91}. We note
that some CVs, as well as accreting compact objects with massive
companions, or qLMXBs with (sub)giant secondaries can have $\log
(F_{\rm X}/F_{\rm O})_{\rm u} \lesssim -1$ as well. We use the $B_X$
band and the $R$ magnitude where possible, or the $V$ magnitude
otherwise, to calculate this ratio. The X-ray flux is obtained from
the best spectral fit. X-ray and optical fluxes are corrected for
absorption using the column density derived from the X-ray spectral
fit ($N_{{\rm H,X}}$).

If we assume that the extinction arises primarily from the Galactic
line-of-sight column density, we can use the three-dimensional dust
model of \citet{drimea03} to derive distances. These are used in turn
to calculate X-ray luminosities $L_{\rm X}$. Distance errors are
calculated by inputing the 1-$\sigma$ errors on $N_{\rm H,X}$ to the
dust model, and are propagated to estimate errors on $L_{\rm X}$. This
approach does not take into account systematic uncertainties in the
\cite{drimea03} model that occur even after applying the ``rescaling''
factors to bring the model closer to the observed COBE/DIRBE far-IR
data, which we have done here. To estimate the potential systematic
errors in our distances, we have compared the $A_V$-versus-distance
curves from Drimmel et al.~with those from \cite{marsea06}, available
for $|\,l\,|<100^{\circ}$. The Marschall et al.~extinction maps have a
higher spatial resolution (15\arcmin~compared to 35\arcmin) and are
derived by comparing 2MASS photometry with a Galactic stellar
population model. We find that the differences can be large, up to a
factor of 2. We note that our assumption that the extinction stems
only from the line-of-sight column density is not necessarily
justified. For sources that are internally absorbed this assumption
can lead to overestimates of the distance, and hence the X-ray
luminosity. Furthermore, for sources that lie outside the disk, this
method can only provide a lower limit to the distance. Cases for which
$N_{\rm H,X}$ is close the asymptotic part of the
extinction-versus-distance curve are marked in Table
\ref{flux-dist-lum}, which summarizes our final classification and
other derived properties.  The errors on $F_{\rm X,u}$ and $\log
(F_{\rm X}/F_{\rm O})_{\rm u}$ in Tables~\ref{spectral-param} and
\ref{flux-dist-lum} are only based on the errors on the net counts and
the magnitude errors, and do not include a contribution from the
uncertainties in the spectral fits. In the remainder we assume $N_{\rm
  H}=(1.79 \times 10^{21}) \times A_V$ cm$^{-2}$
\citep{predschm95}. Absolute magnitudes and bolometric luminosities as
a function of spectral type are taken from \cite{ostlcarr07} unless
mentioned otherewise.

\begin{table}
\caption{Classification and derived properties for the bright sources\label{flux-dist-lum}}
\begin{center}
\begin{tabular}{ccccl}
\hline
\hline
CBS & $d$ & $\log L_{\rm X}$ & $\log (F_{\rm X}/F_{\rm O})_{\rm u}$ & comment \\
    & kpc      &               & &  \\
\hline
\multicolumn{5}{c}{Stars} \\
\hline
3 & 1$\pm$0.4 & $31.1\pm0.3$ & $-2.50\pm0.03$ & \ldots \\
4 &1.8$\pm$0.3 & 31.4$\pm$0.2 & $-0.93\pm0.05$ & \ldots \\
5 & 2.8 & 32.88$\pm$0.02 & $-4.03\pm0.03$ & \ldots \\
6 & 0.9$\pm$0.7 & 31.2$^{+0.5}_{-1.6}$ & $<-0.5$\tablenotemark{c}& \ldots \\
8  & 3.4$\pm$0.9 & 32.2$\pm$0.3 & $-2.40\pm0.05$ & \ldots \\
10 & $<0.5$ & $<31.2$ & $-1.84\pm0.04$ & \ldots\\
11 & $<0.6$ & $<30.1$ & $-2.39\pm0.05$ & \ldots\\
12 & $<1.3$ & $<30.8$ & $-2.95\pm0.05$ & \ldots\\
14 & $<0.5$ & $<32$\tablenotemark{e} & $1.64\pm0.04$\tablenotemark{e} & flare star \\
19& 3.0$\pm$0.8 & 31.8$\pm$0.3 & $-2.20\pm0.05$ & \ldots\\
20& $<3$ & $<31.5$ & $-2.50\pm0.05$ & \ldots\\
21& $<0.2$ & $<30.0$ & $-2.36\pm0.04$ & \ldots\\
\hline
\multicolumn{5}{c}{Accreting binaries} \\
\hline
7  & 1.4$\pm$0.1\tablenotemark{f} & 32.05$\pm$0.08\tablenotemark{f} & 1.31$\pm$0.07 & CV\\
17 & 4.9$\pm$0.8 & 33.2$\pm$0.2 &  $-1.58\pm0.04$\tablenotemark{d} & symbiotic?\\
\hline
\multicolumn{5}{c}{AGN} \\
\hline
2 & $>2.4$\tablenotemark{a} & $>31.8$\tablenotemark{a} & $-$0.12$\pm$0.05 & \ldots \\   
\hline
\multicolumn{5}{c}{AGN or accreting binaries} \\
\hline
1 & $>7.2$\tablenotemark{a} & $>32.8$\tablenotemark{a} & 0.20$\pm$0.05 & \ldots  \\  
9 & $>16$\tablenotemark{a} & $>33.8$\tablenotemark{a} & $-$2.34$\pm$0.05 & \ldots \\       
13 & $>11$\tablenotemark{a} & $>33.7$\tablenotemark{a}         & $> -7.6$\tablenotemark{b} & \ldots \\
15 & $>23$\tablenotemark{a} & $>34.4$\tablenotemark{a}         & $> 0.3$\tablenotemark{b} & \ldots \\
16 & $>4.4$                  & $>32.5$         & $> 1.1$\tablenotemark{b} & \ldots \\ 
18 & 5.8$\pm$2.4             & 33.0$^{+0.3}_{-0.5}$ & $> -0.9$\tablenotemark{b} & \ldots \\
\hline
\end{tabular}

\tablenotetext{1}{$N_{\rm H,X}$ exceeds the integrated Galactic column
  density ($N_{\rm H,Gal}$) in the direction of the source according
  to \cite{drimea03} by more than 1 $\sigma$. We set the lower limit
  on $d$ to the distance where the extinction curve reaches its
  asymptotic value of $A_V$.}  \tablenotetext{2}{These sources lack
  candidate optical counterparts. We assume $R\gtrsim24$.}
\tablenotetext{3}{This star is saturated in our images. We assume
  $R\lesssim17$.}  \tablenotetext{4}{We use $N_{\rm H,X}$ to deredden
  the X-ray flux, and $N_{\rm H}=3.25\times10^{21}$ to correct the
  optical flux. See text in \S\ref{results_cbs17}.}
\tablenotetext{5}{After correcting for pileup. The uncertainty in the
  pileup correction is not included. See text in \S~\ref{x-sample}.}
\tablenotetext{6}{Distance and luminosity are based on $N_{\rm
    H,X}$ obtained from fitting the {\em XMM-Newton} spectrum.}

\tablecomments{Columns: 1) source number; 2) spectroscopic distance
  for CBS 5 and 17; for the remaining sources this is the distance
  derived from the \cite{drimea03} model assuming that $N_{\rm H,X}$
  arises {\em primarily} from the Galactic line-of-sight
  extinction, but see \S\ref{source_class} for caveats regarding
    these distances; 3) X-ray luminosity (0.3--8 keV); 4) ratio of
  dereddened X-ray and $R$-band flux or $V$-band flux (for CBS 10 and
  21), where only the errors on the X-ray net counts and the
    optical magnitudes are included in the uncertainties.  Maximum
  (minimum) values on distance and $L_{\rm X}$ are 1-$\sigma$ upper
  (lower) limits that result from the 1-$\sigma$ upper (lower) limit
  on $N_{\rm H,X}$. }
\end{center}
\end{table}

\subsection{Cross-correlation with other X-ray catalogs} \label{sec_xcat}
Many of our sources have potential counterparts in the {\em
  XMM-Newton} Serendipitous Source Catalog DR3 \citep{watsea09}. We
give the names of sources within 3\arcsec~from the {\em Chandra}
position in column 13 of Table~\ref{x-sample-table}. For CBS 7, 14,
and 17 we have done a further analysis of the {\em XMM-Newton}
observations that include these source positions. We have
cross-correlated our sample with the {\em ASCA} Galactic Plane Survey
source catalog \citep{sugiea01} but found no matches within the
1\arcmin~positional uncertainty of the {\em ASCA} sources.

\section{Results}\label{results}

\subsection{Accreting binaries}

There is compelling evidence that CBS 7 and 17 are Galactic binaries
whose X-rays are powered by accretion onto a compact object, most
likely a white dwarf.

\subsubsection{CBS 7: a cataclysmic variable} \label{results_cv}

CBS 7 has a hard spectrum that is best fit with a power law with
$\Gamma=1.0 \pm 0.1$; fits with thermal models only place a lower
limit to the plasma temperature of $kT\gtrsim50$ keV. Photometry of
the optical source inside the error circle gives $\log (F_X/F_R)_{\rm
  u} = 1.3\pm0.1$. The $VRI$ colors of this source are blue compared
to the bulk of the surrounding stars, and there is marginal evidence
for excess H$\alpha$ emission with H$\alpha - R = -0.5\pm0.2$. We
present optical follow-up spectra that confirm the H$\alpha$ emission
lines in CBS 7 elsewhere \citep{serveaastroph11}. These
characteristics are typical for accreting binaries with compact
objects. Using the derived $N_{\rm H,X}$, we find a distance $d =
1.2\pm0.5$ kpc, and an absolute magnitude $M_V\approx10.0$. We
conclude that CBS 7 has a low-mass, unevolved donor and is likely a CV
rather than a quiescent wind-accreting Be X-ray binary. Sources
belonging the latter category can have similary hard X-ray spectra but
their massive donors are much brighter in the optical.

\begin{figure}
\centering
\includegraphics[width=7cm,bb=0 23 340 285,clip]{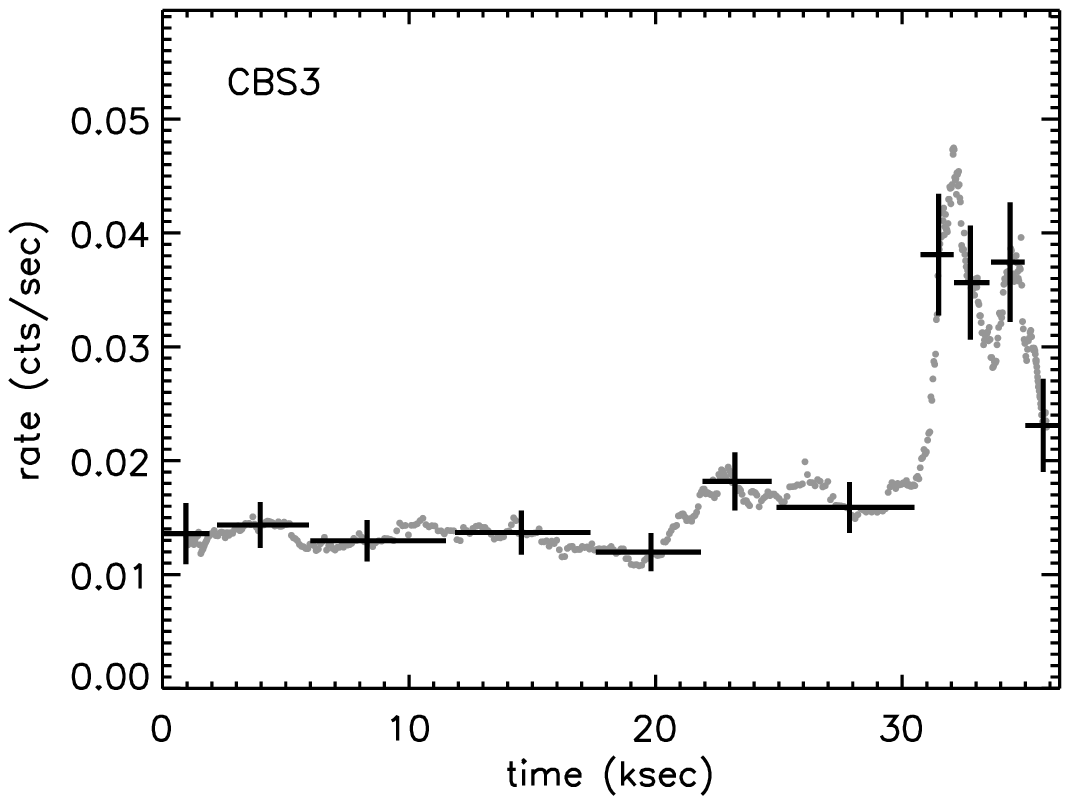}\\
\includegraphics[width=7cm,bb=0 0 340 240,clip]{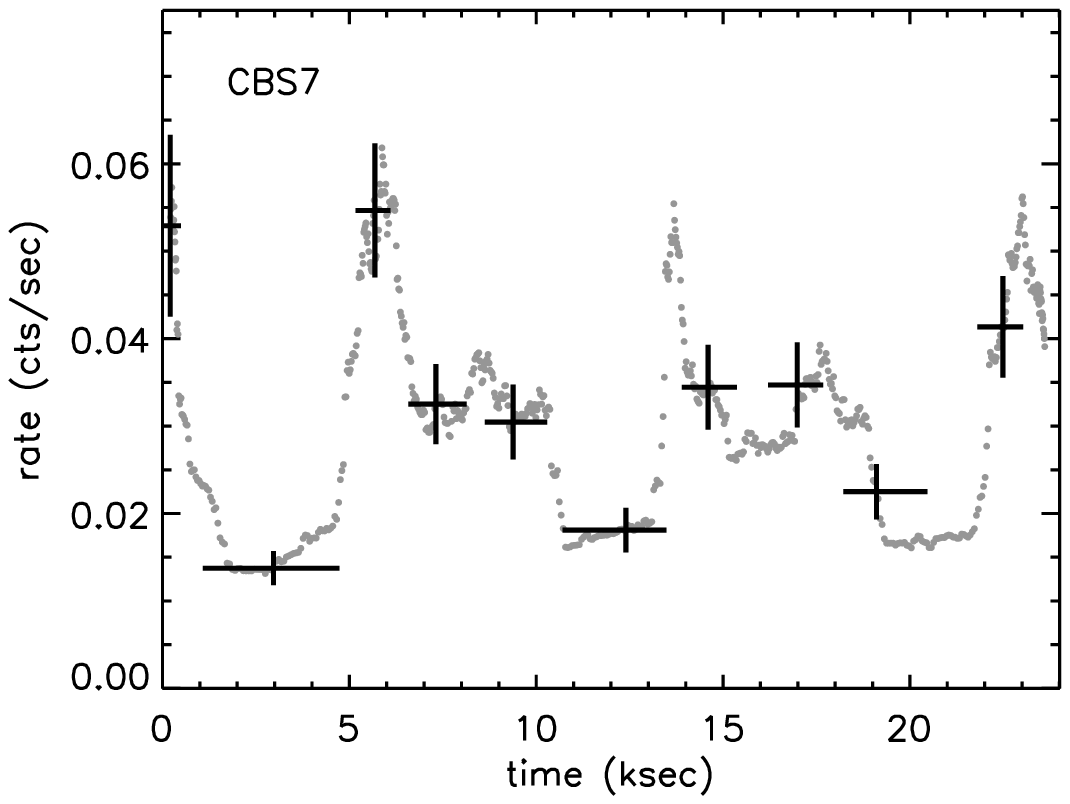}
\caption{Background-corrected {\em Chandra} light curves of CBS 7 and
  CBS 3 (0.3--8.0 keV). The points are average count rates in a sliding
  window with a width that is adjusted to include 50 counts. Error
  bars are shown for a few representative points where the horizontal
  error bar marks the bin width.}
\label{xlc_cbs3_7}
\end{figure}

\begin{figure*}
\center
\includegraphics[width=5.3cm]{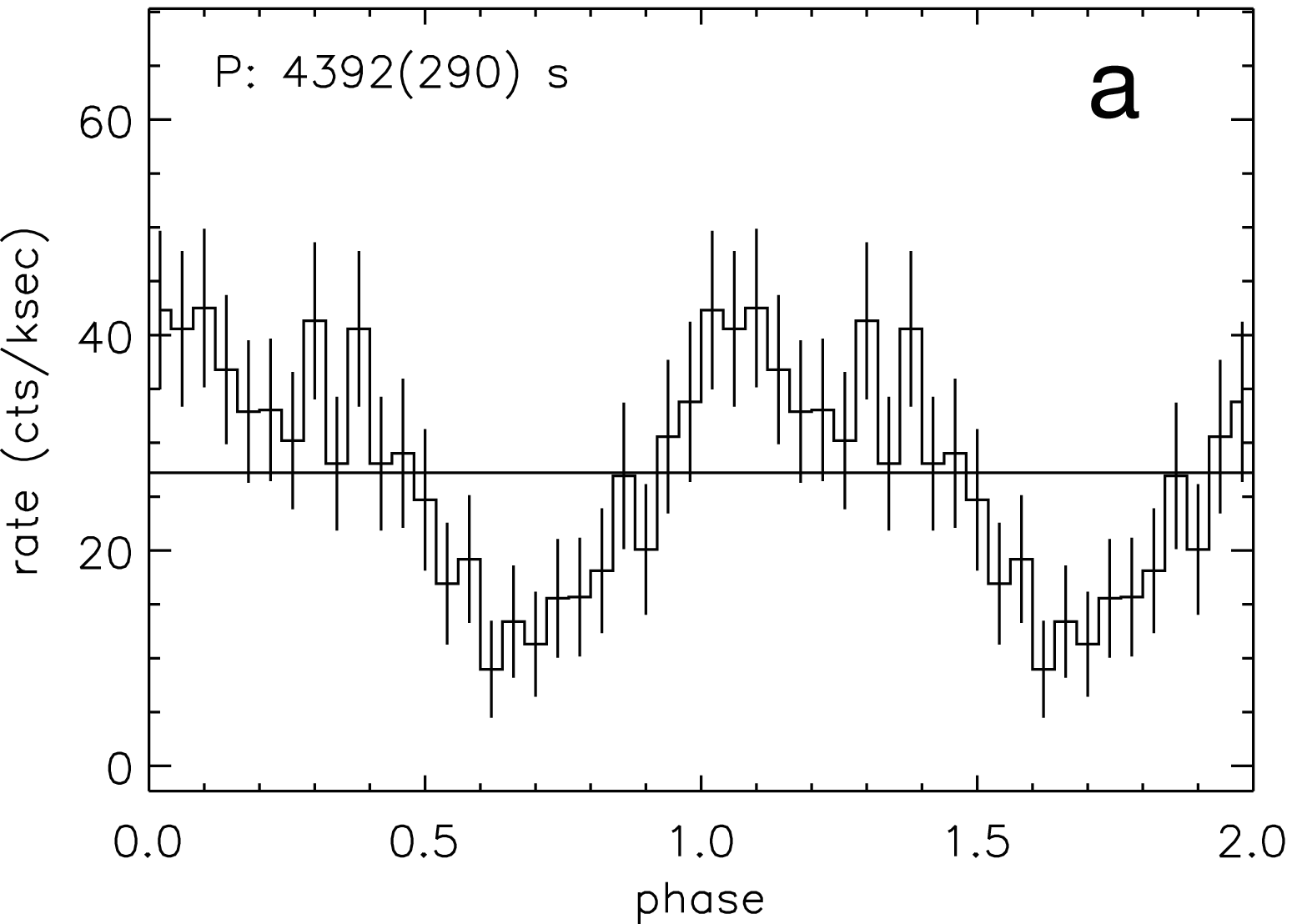}
\includegraphics[width=5.3cm]{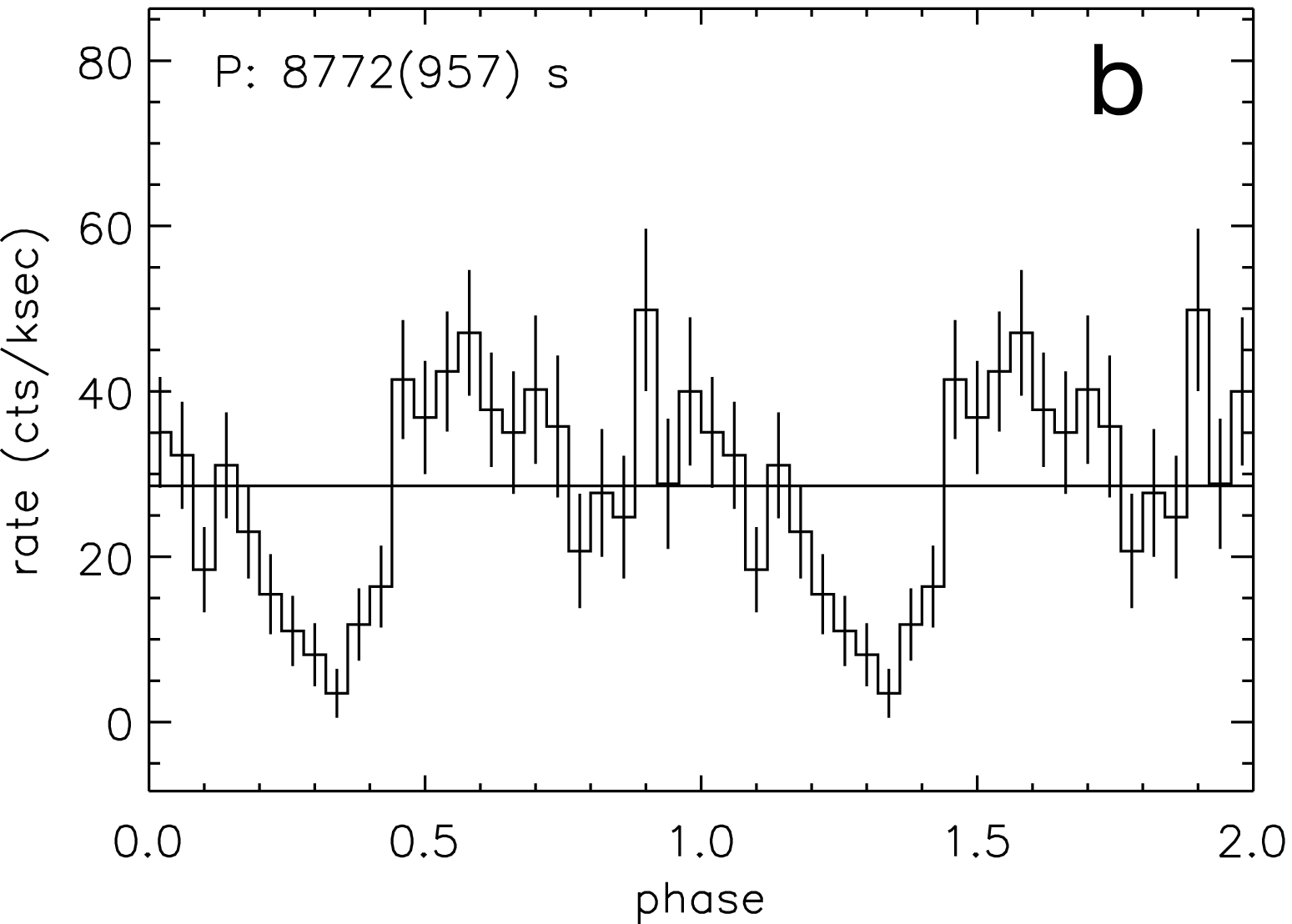}
\includegraphics[width=5.3cm]{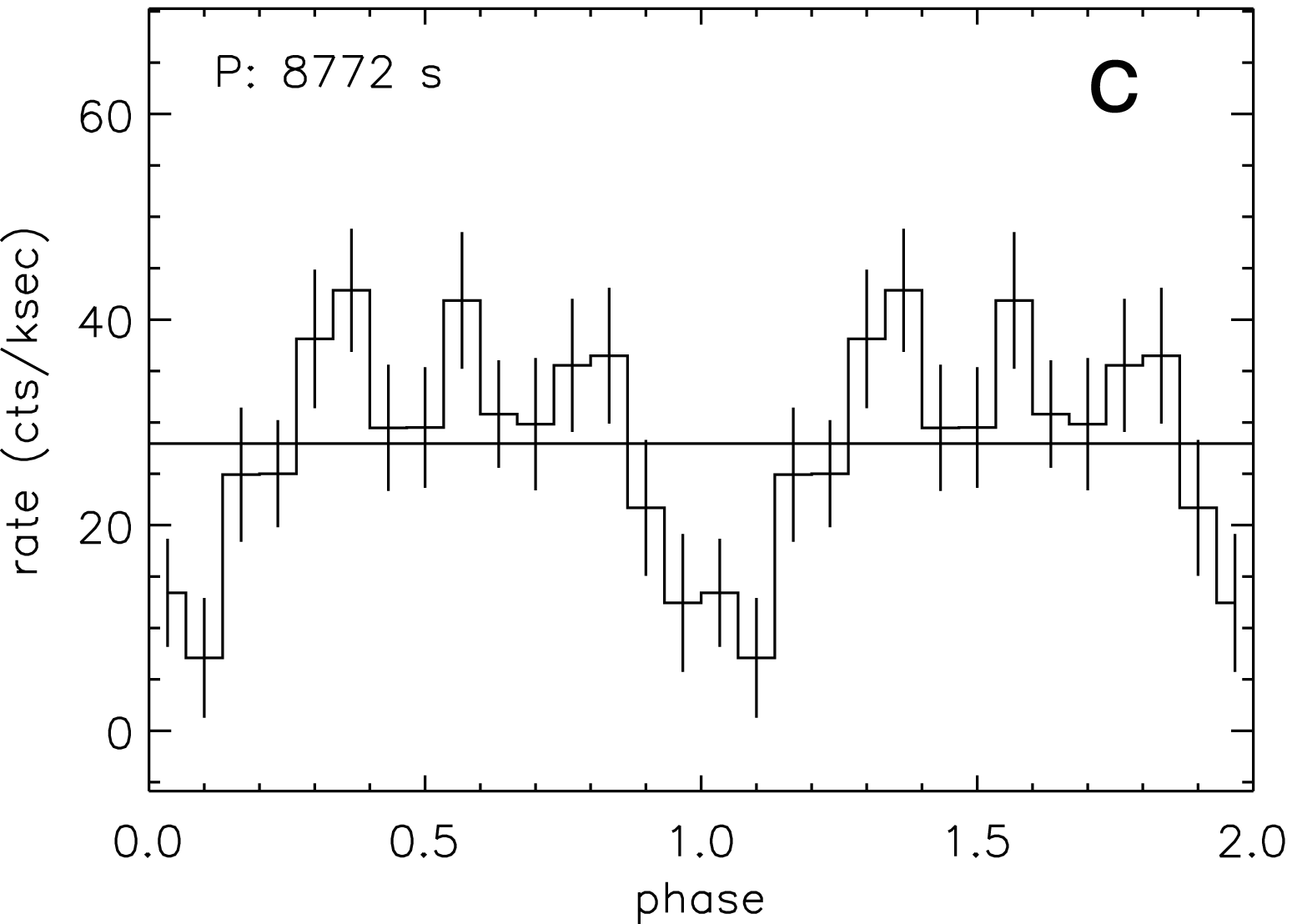}
\caption{{\em Left and middle:} {\em Chandra} light curves for CBS 7
  (0.3--8 keV) folded on the periods $P_1$ and $P_2$ found by the
  Lomb-Scargle analysis. {\em Right:} {\em XMM-Newton} EPIC-PN light
  curve (0.3--8 keV) folded on $P_2$. Horizontal lines mark the
  average count rate. \label{xlc_cbs7}}
\end{figure*}

The X-ray light curve of CBS 7 is clearly variable, with recurring
deep and shallow dips (Fig.~\ref{xlc_cbs3_7}). We ran a Lomb-Scargle
period-search algorithm \citep{scar} on the barycenter-corrected
photon arrival times. \cite{hongeaastroph11} give details of the
timing analysis. The power-density spectrum between 20 s and 20\,000 s
shows peaks with $>$99\% significance at $P_1 = 4392 \pm 290$ s and
$P_2 = 8772 \pm 957$ s. The longer period corresponds to the spacing
of the deep dips, while the shorter one is the spacing between the
deep and shallow dips. Figs.~\ref{xlc_cbs7}a and b show the {\em
  Chandra} light curves folded on $P_1$ and $P_2$.  We examined the
time-resolved spectral parameters using quantile analysis, but see no
indication for variations correlated with count rate. This could be
partly due to poor statistics.

CBS 7 is also serendipitously included in an {\em XMM-Newton}
observation taken on 2003, Aug 14 (ObsID 0152780201; 81 ks). We
retrieved the data from the archive and performed further processing
with SAS 10.0 following instructions in the {\em XMM-Newton} ABC
Guide\footnote{http://heasarc.gsfc.nasa.gov/docs/xmm/abc/} and the SAS
Threads\footnote{http://xmm.esac.esa.int/sas/current/documentation/threads/}.
After filtering out time intervals with background flares, $\sim$37 ks
(EPIC-PN), 59.2 ks (EPIC-MOS1), and 62.1 ks (EPIC-MOS2) of exposure
time remains. Periods consistent with the values of $P_1$ and $P_2$
show up as significant peaks in the periodograms of the PN light
curves. For the MOS1 and MOS2 data only the shorter period is
significantly detected, but the folded light curves do not show
convincing variability. On the other hand, folding the PN and MOS
count rates on $P_2$ and $P_2/2$ produces light curves that are
qualitatively similar to the {\em Chandra} light curves and give a
smoother result than the periods derived from the {\em XMM-Newton}
data (Fig.~\ref{xlc_cbs7}c).

To investigate the spectrum, we focused on the data from the PN camera
given its larger effective area compared to the MOS cameras and {\em
  Chandra} ACIS at higher energies. We extracted counts from a
20\arcsec~source aperture, and corrected for background using a nearby
source-free region on the same chip. Fitting a thermal-bremsstrahlung
model to the PN spectrum, grouped to have at least 20 counts per bin,
sets a lower limit to the temperature of $kT>30$ keV. A power-law fit
gives parameters that are consistent with those in
Table~\ref{spectral-param}, viz.~$\Gamma=1.30\pm0.06$ and $N_{\rm
  H,X}=(2.4\pm0.2)\times10^{21}$ cm$^{-2}$ ($\chi^2_{\nu}=1.07$, 91
d.o.f.). The {\em XMM}-derived value for $N_{\rm H,X}$ is thus better
constrained than the value obtained by fitting the {\em Chandra}
spectrum, and gives a distance of $1.4\pm0.1$ kpc; this is the value
we include in Table~\ref{flux-dist-lum}. Note however the caveat in
\S\ref{disc_cbs7} regarding our distance estimate for this source.
The residuals show a systematic excess between 6 and 7 keV compared to
this model, which prompted us to add a gaussian-shaped emission line
at a fixed position of 6.4 keV or 6.7 keV to mimic an Fe K$\alpha$
fluorescent or He-like emission line. While the former does not give
sensible line parameters, including a 6.7 keV line removes the
systematic residuals. In the latter case, we find a marginal (95\%)
significance for the presence of the line using the method of Bayesian
posterior predictive probability values \citep[e.g.~][]{protea02}. The
spectrum can be described by $\Gamma=1.36\pm0.07$, $N_{\rm
  H,X}=(2.6\pm0.3)\times10^{21}$ cm$^{-2}$, and a FWHM of 0.6$\pm$0.2
keV ($\chi^2_{\nu}=0.98$, 89 d.o.f.). The unabsorbed flux of $F_{\rm
  X,u}=6.3\times10^{-13}$ erg s$^{-1}$ cm$^{-2}$ (0.3--8 keV) is about
25\% higher than found with {\em Chandra}.

\subsubsection{CBS 17: a symbiotic binary?} \label{results_cbs17}

\begin{figure}
\center
\includegraphics[width=8.5cm,bb=14 0 415 283]{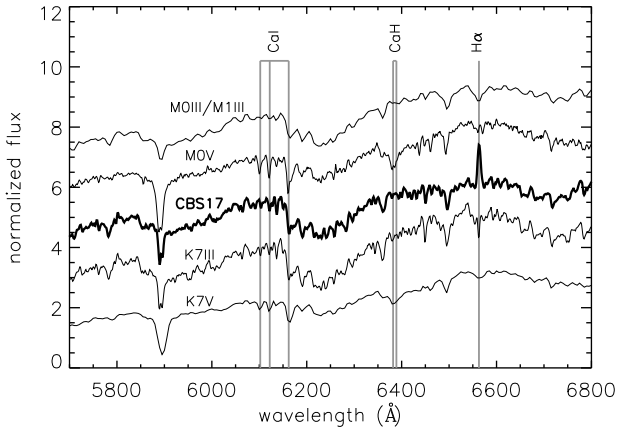}
\caption{The optical spectrum of the likely counterpart to CBS 17
  ({\em middle; thick line}) is compared to template spectra of dwarfs
  and giants of similar spectral type \citep{jacoea84,silvcorn92} that
  are reddened with $N_{\rm H}=3.5\times10^{21}$ cm$^{-2}$. Our star
  clearly shows H$\alpha$ emission. The vertical lines mark CaI (6102,
  6122, 6162~\AA) and CaH (6382, 6389~\AA) lines that are prominent in
  dwarfs but weak or absent in giants and in CBS 17.\label{fig_cbs17}}
\end{figure}

CBS 17 is detected $\lesssim$10 pixels away from the chip edge, so we
use caution when considering its properties derived from the {\em
  Chandra} data. The hard power-law spectrum ($\Gamma = 0.8 \pm 0.2$;
or $kT>43$ keV for a thermal model) suggests this source is
accretion-powered. TiO absorption bands (5847--6058 \AA, 6080--6390
\AA, 6651--6852 \AA) are clearly visible in our Hydra spectrum of the
candidate counterpart, and their strength points at a late-K or
early-M spectral type. The absence of the gravity-sensitive CaH lines
around 6382 and 6389 \AA\, indicate this star is a giant
(Fig.~\ref{fig_cbs17}). H$\alpha$ is seen in emission, which supports
the true association with the X-ray source. On these grounds we
conclude that CBS 17 is likely a symbiotic binary where an evolved
late-type star transfers mass to a hotter companion, in many cases a
white dwarf. Many ``canonical'' symbiotics are thought of as
wind-accreting systems as the strong wind from the evolved companion
manifests itself through nebular emission lines excited by the
high-energy photons created by the accretion process. We do not see
such prominent emission lines in CBS 17. Mass transfer would therefore
have to take place through Roche-lobe overflow.

The observed nIR colors of CBS 17 are only consistent with the
spectral classification if the extinction is lower ($N_{\rm H} \approx
3 \times 10^{21}$ cm$^{-2}$) than the one derived from X-ray spectral
fitting (Fig.~\ref{fig_ccd}b). A lower value for $N_{\rm H}$ of about
$3.5 \times 10^{21}$ cm$^{-2}$ is also suggested by comparing the
continuum slope of the Hydra spectrum to artificially reddened
template spectra. Formally the error bar on $N_{\rm H,X}$ allows such
low values (the difference is $\sim$2.3\,$\sigma$), but the
discrepancy can also indicate that there is material inside the system
obscuring the X-ray emitting region but not the late-type
giant. Therefore, we do not use $N_{\rm H,X}$ to estimate a distance,
but adopt $N_{\rm H}=(3.25\pm0.25) \times 10^{21}$ cm$^{-2}$ to
calculate a spectroscopic distance based on the allowed range of
spectral type.  A K7\,III giant has $M_V=+0.4$, while an M1\,III giant
has $M_V=-0.2$, resulting in $d=4.1-5.6$ kpc, and $\log L_{\rm
  X}=33.2\pm0.2$.

\nocite{angewhit03} CBS 17 is included in three {\em XMM-Newton}
observations.  It is commented on by Angelini \& White (2003, AW03;
their source 1) as a serendipitous detection and possible AGN that
stands out in a 6.2--6.8 keV Fe-band image of the first of these
observations. \cite{kaarea06} also list CBS 17 among the serendipitous
detections in the first and second {\em XMM-Newton} observation, as
well as in the {\em Chandra} observation analyzed here.  Their
reported fluxes indicate long-term variability but are based on the
assumption (for all their sources) of a power-law spectrum with
$\Gamma=1.5$ and $N_{\rm H} = 3.1 \times 10^{21}$ cm$^{-2}$. This
motivated us to do a more detailed
analysis. Table~\ref{tab-cbs17-xmmobs} lists all X-ray observations
for CBS 17.

\begin{table}
\begin{center}
\caption{X-ray observations of CBS 17 used in our analysis} \label{tab-cbs17-xmmobs}
\begin{tabular}{lllll}
\hline
\hline
epoch       & telescope/instrument     & ObsID       & Date Obs & T$_{\rm exp}$\tablenotemark{a}\\ 
            &                          &             &         & ks          \\
\hline
1           & {\em XMM-Newton}/EPIC-PN & 0032940101  & 2001-03-08 & 15.4 \\
2           & {\em Chandra}/ACIS-S     & 04586       & 2004-06-25 & 44.1 \\
3           & {\em XMM-Newton}/EPIC-PN & 0203750101  & 2004-09-18 & 42.7 \\
4           & {\em XMM-Newton}/EPIC-PN & 0500540101  & 2008-03-15 & 44.6 \\
\hline
\end{tabular}

\tablenotetext{1}{GTI exposure time}
\end{center}
\end{table}

We extracted spectra following the same procedures as for CBS 7, and
restricted the analysis to the EPIC-PN data. First we fitted each
spectrum individually. The results, summarized in the upper part of
Table~\ref{xmmspectral-param}, indicate spectral and/or flux
variability but the errors are large. Therefore we also tried fitting
all spectra simultaneously, forcing various combinations of parameters
to be the same for each epoch. Keeping $N_{\rm H,X}$ and the power-law
slope and normalization the same gives an unacceptable fit with
$\chi^2_{\nu}=1.81$ and 278 d.o.f. for the overall fit, which confirms
the variability. Allowing only $N_{\rm H,X}$ to vary for each epoch
does not result in a good fit either (overall $\chi^2_\nu=1.69$, 275
d.o.f.). The same is true if we fix $N_{\rm H,X}$ and $\Gamma$, but
allow the normalization to vary; in this case an unsatisfactory fit is
found for epoch 1 and 2. If we let $\Gamma$ and the normalization vary
and keep $N_{\rm H,X}$ the same, a good fit is found for each epoch
(Table~\ref{xmmspectral-param}, bottom). In this case we see small
variations of $\Gamma$, and the intrinsic flux varies up to
$\sim$60\%. The largest flux change is seen between two {\em
  XMM-Newton} observations (epoch 1 and 3), so this finding is
unaffected by the fact that CBS 17 lies close to the chip edge in the
{\em Chandra} observation.

\begin{figure}
\centering
\includegraphics[width=8cm,bb=0 25 420 420]{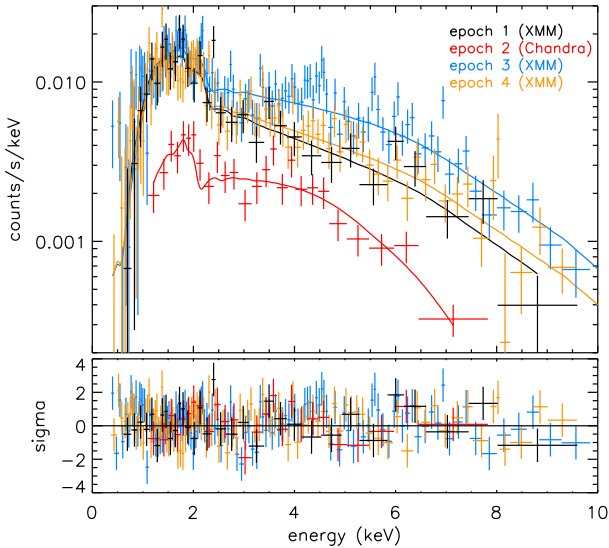}
\caption{{\em Chandra} and {\em XMM-Newton} spectra of CBS 17 are
  shown together with the best fitting model to the data. Solid lines
  are the model fits for an absorbed power law with the column density
  $N_{\rm H,X}$ kept the same for each epoch. The model parameters are
  listed in the bottom part of Table~\ref{xmmspectral-param}. {\em See
    the electronic edition of the Journal for a color version of this
    figure.}}
\label{fig-cbs17-xspec}
\end{figure}

Fig.~\ref{fig-cbs17-xspec} shows the data and best model fits for
fixed $N_{\rm H}$. The spectrum from epoch 1 shows small positive
residuals between 6 and 7 keV that could correspond to the Fe K line
reported by AW03. The good fits achieved with models without an Fe K
line do not warrant adding such an emission component. When we {\em
  do} include a gaussian-shaped emission line, we find parameter
values that are consistent with the results of AW03,
viz.~6.24$\pm$0.15 keV and 0.66$\pm$0.24 keV for the line center and
width, $\Gamma=1.7\pm0.2$, $N_{\rm H,X}=(0.9 \pm 0.2) \times 10^{22}$
cm$^{-2}$, and $F_{\rm X,u}=4.8\times10^{-13}$ erg s$^{-1}$ cm$^{-2}$
($\chi^2_{\nu}=0.91$ for 39 d.o.f.). Here we re-grouped the spectrum
to $\geq$15 (instead of 20) counts per bin for a slightly better
resolution. Only the spectrum from epoch 1, during which the source
was at its softest, shows a hint of a line. The absence of this line
in the {\em Chandra} data could be due to the poorer sensitivity at
higher energies.

Follow-up studies are needed to uncover the nature of CBS 17.  With
high-resolution optical spectroscopy the binary status can be
established, and the orbital period and companion masses can be
constrained. To test if the companion is (close to) Roche-lobe
filling, one can look for the signature of a distorted companion in
the optical/nIR light curves.

\begin{table}
\begin{center}
\caption{X-ray spectral fits for CBS 17 \label{xmmspectral-param}} 
\begin{tabular}{ccccc}
\hline
\hline
epoch & $\Gamma$ & $N_{\rm H,X}$        & $F_{\rm X,u} \times 10^{-13}$ &  $\chi^2_{\nu}$/d.o.f  \\
      &          & 10$^{22}$ cm$^{-2}$  & erg s$^{-1}$ cm$^{-2}$        &                        \\
\hline				                                     
\multicolumn{5}{l}{{\em each spectrum fit individually}}\\
1   & 1.42$\pm$0.16 & 0.76$\pm$0.16     & 4.2                          &  1.02/30                \\
2   & 1.0$\pm$0.3   & 0.8$\pm$0.2       & 5.2                          &  0.91/23              \\
3   & 0.87$\pm$0.07 & 0.70$\pm$0.10     & 6.6                          &  1.08/120              \\
4   & 1.19$\pm$0.08 & 0.66$\pm$0.10     & 4.1                          &  0.93/96         \\
\hline
\multicolumn{5}{l}{{\em same $N_{\rm H,X}$ for each epoch}}\\
1   & 1.38$\pm$0.10 & 0.71$\pm$0.06   & 4.1                          &  1.00/272               \\
2   & 0.64$\pm$0.10 & .          & 5.0                          &  .       \\
3   & 0.87$\pm$0.06 & .          & 6.6                          &  .             \\
4   & 1.23$\pm$0.07 & .          & 4.2                          &  .             \\
\hline
\end{tabular}
\tablecomments{A ``.'' indicates that the parameter was forced to be the same for each epoch.}
\end{center}
\end{table}

\subsection{Stellar coronal X-ray sources}\label{results_stars}

We classify 9 sources (CBS 3, 4, 5, 6, 10, 11, 12, 20, 21) as stars or
active binaries based on the following criteria: $\log (F_{\rm
  X}/F_{\rm O})_{\rm u} \lesssim -1$ \citep[][and references
  therein]{koenea08}, the X-ray spectrum is well fit by a thermal
model, and the corresponding plasma temperature $kT$ is a few keV at
most. Their properties are described in \S\ref{results_ltstars}
(late-type stars) and \S\ref{results_etstars} (early-type star). Three
exceptional cases---CBS 19, 8, and 14---are discussed in
\S\ref{results_pms}--\ref{results_ultracool}.

\subsubsection{Normal and active late-type stars}\label{results_ltstars}

Our optical spectra show that CBS 3, 4, 11, 12, 20 and 21 are G or K
stars (Table~\ref{opt-sample-table}). We have not obtained spectra for
the candidate counterparts to CBS 6 and 10, yet. Considering their
dereddened 2MASS colors, they are probably late-type stars as well,
with spectral types around early G and early M, respectively
(Fig.~\ref{fig_ccd}b). The X-rays of late-type stars are emitted by a
hot coronal plasma, which is confined by magnetic fields that are
generated by a solar-like dynamo operating in the convective outer
layers. See \cite{gued04} for a review.

\nocite{dempea,dempea97} The X-ray derived extinction values for these
eight stars are all relatively low, and consequently the Drimmel
extinction-versus-distance curves place them nearby. Our estimates or
upper limits on $L_X$ are consistent with X-ray luminosities of nearby
F--K stars \citep[$\log L_{\rm 0.1-2.4~keV} \approx
  26.5-29.5$;][]{schmlief04} and active binaries ($\log L_{\rm
  0.1-2.4~keV} \approx 28-32$; Dempsey et al.~1993, 1997) as measured
by {\em ROSAT}, and of active binaries in open clusters \citep[$\log
  L_{\rm 0.3-7~keV} \approx 28-31$;][]{vdbergea04} as measured by {\em
  Chandra}. If we adopt the $N_{\rm H,X}$-derived distances to
calculate absolute magnitudes $M_V$ or $M_R$, we find that CBS 4, 10,
11, and 21 are likely main-sequence stars or BY\,Dra-type active
binaries, and CBS 3 is likely a subgiant or an RS\,CVn-type active
binary. CBS 12 and 20 are dwarfs or subgiants, either single or in an
active binary. While one should be wary when using our $N_{\rm
  H,X}$-derived distance estimates, we note that these sources are
relatively nearby, and that non-flaring active stars or binaries are
often not internally absorbed.  Low-resolution stellar X-ray spectra
are often found to be better fit with a sub-solar coronal metal
abundance $Z$ than with a solar abundance \citep{gued04}. We also see
this in the results of our fits to the spectra of CBS 3, 6, 10, 11,
12, 20, and 21.

With $kT\approx4$ keV and $\log (F_{\rm X}/F_R)_{\rm u} \approx -0.9$,
CBS 4 appears to be more active than most of the sources in this
category, which have $kT\lesssim1$ keV. Here we could be seeing the
effect of a coronal flare during the first $\sim$20 ks of the
observation. To study temporal variations in the energy spectrum of
CBS 4, we resort to quantile analysis as there are too few counts to
fit, and compare, spectra extracted from different sub-intervals of
the exposure. In quantile analysis, the energy values that divide the
energy distribution of the photons in certain fixed fractions are used
as spectral diagnostics. This approach is more powerful than measuring
the number of counts in fixed energy bands (as is done when using
hardness ratios), as the errors on the diagnostic are less sensitive
to the underlying spectral shape. Following \cite{hongschlea04} we
choose the median energy ($E_{50}$), and the 25\% and 75\% quartiles
($E_{25}$, $E_{75}$) to characterize the spectra. By comparing the
observed quantiles with those expected for a spectral model of choice,
the X-ray spectral parameters and $N_{\rm H,X}$ can be constrained. As
expected for a coronal flare, the spectrum of CBS 4 is harder when it
is bright compared to quiescence, where the spectrum can be described
by $\sim$1 keV thermal plasma. For intermediate count rates ($\sim$10
cts ks$^{-1}$) the source moves down in the quantile diagram which is
the signature of two different temperatures contributing more or less
equally.  Flaring has been frequently observed in M dwarfs, but also
in earlier-type dwarfs (see \citealt{gued04} and references
therein). What is puzzling is the trend of spectral hardening towards
the end of the observation, when the count rate is at its lowest.

CBS 3 is also an X-ray variable, showing a flare-like event near the
end of the exposure (Fig.~\ref{xlc_cbs3_7}). We see no indication for
a change in its X-ray spectrum during the flare.

CBS 12 is potentially a long-term variable. Besides in the observation
we analyze here, it is detected in ObsID 757 (2000 Aug 14, 13.4 ks
GTI) with a significantly higher count rate: 9.8$\pm$1.2 versus
4.3$\pm$0.2 cts ks$^{-1}$. The energy quantiles in the two ObsIDs are
consistent.

\begin{figure*}
\centering
\includegraphics[width=18cm]{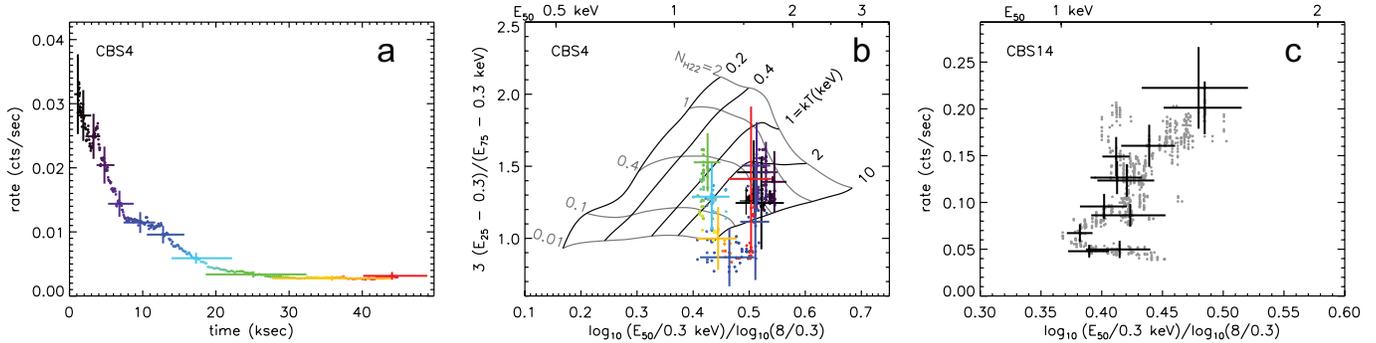}
\caption{{\em Left:} Background-corrected {\em Chandra} light curve of
  CBS 4 (0.3--8.0 keV). See the caption of Fig.~\ref{xlc_cbs3_7} for a
  description of the binning. {\em Middle:} Time-resolved quantile
  diagram (0.3--8 keV) for CBS 4. Each point corresponds to the bin of
  the same color in the left panel. The points are plotted on a
  thermal-bremsstrahlung grid, where the black lines represent a
  constant temperature $kT$ and the gray lines represent a constant
  column density $N_{\rm H}$ (in units of 10$^{22}$ cm$^{-2}$).
    By comparing the observed quantiles with these grid lines, one can
    constrain the spectral parameters for this assumption of spectral
    model. A clear concentration towards two temperatures is
  seen. Note that the spectrum appears to become harder again in the
  last two time bins when the count rate is low. {\em Right:} The
  0.3--8 keV count rate for CBS 14 shows an overall positive
  correlation with spectral hardness. Events from the piled-up core
  are excluded. {\em See the electronic edition of the Journal for a
    color version of this figure.}}
\label{xlc_cbs4_14}
\end{figure*}

\subsubsection{CBS 5: an early-type star}\label{results_etstars}

The bright star that is matched to CBS 5 is HD\,97434. The O7.5III
classification by \cite{walb73} is in better agreement with our Hydra
spectrum than other classifications found in the literature.  Using
its colors and spectral type, \cite{vazqfein90} derive
$A_{V}=1.5\pm0.2$ and $M_V=-5.7\pm0.15$, where the errors were
estimated by us (since none were given by Vazquez \& Feinstein) and
set to reasonable values based on the information provided in their
study, and the compilation of absolute magnitudes of O stars in
\cite{vaccea96}. This gives a distance to CBS 5 of $d=2.8\pm0.4$
kpc. The X-rays from single early-type stars are believed to arise
from shock-heated plasma formed by instabilities in the strong winds
emanating from such massive stars. The spectra can be described by the
combination of two thermal components with $kT\approx0.3$ and 0.7--1
keV \citep{guednaze09}. This agrees with our results. We find $\log
(F_{\rm X}/F_R)_{\rm u} \approx -4.0$, and $\log (L_{\rm X}/L_{{\rm
    bol})} \approx -6.3$ (for $d=2.8$ kpc), also typical for O stars.

\cite{bhatea10} analyzed {\em XMM-Newton} data of HD 97434 taken two
months before the {\em Chandra} observation. To facilitate comparison,
we fit our spectrum with their adopted model, i.e.~a 2$T$ APEC thermal
plasma model ({\em xsapec}). Our data are not good enough to constrain
$Z$. Setting $Z/Z_{\odot}\equiv0.21$, i.e.~the value found by Bhatt et
al, we find $N_{\rm H,X}$ $=(3\pm1) \times 10^{21}$ cm$^{-2}$,
$kT_1=0.24\pm0.05$ keV, and $kT_2=0.6\pm0.1$ keV, suggesting no change
in these parameters between the epochs. Bhatt et al.~assume that HD
97434 is a member of the open cluster Trumpler 18 located at
1.55$\pm$0.15 kpc \citep{vazqfein90}. However, the spectroscopic
distance (see above) places it behind the cluster by
$\sim3\sigma$. When adopting a common distance, the {\em Chandra}
luminosity is 2--3 times lower than the {\em XMM-Newton} value,
indicating long-term variability.

\subsubsection{CBS 19: a pre-main sequence star}\label{results_pms}

The temperature derived for CBS 19 ($kT=4^{+2}_{-1}$ keV) indicates a
high level of activity. Our Hydra spectrum of the candidate
counterpart shows a clear H$\alpha$ emission line with
EW=$-$6.0$\pm$0.8\,\AA. At the poor signal-to-noise ratio of our
spectrum, the continuum looks featureless except for the Na\,I~D
doublet around 5890\,\AA, which could be interstellar in nature. This
makes spectral classification challenging. At first sight, the optical
and nIR colors of the candidate counterpart seem to disagree
(Fig.~\ref{fig_ccd}). The former suggest a spectral type of late-F to
early-M, whereas the latter point at a late-M star. Such cool M stars
show prominent TiO absorption bands in their optical spectra but these
are notably absent in CBS 19.

One explanation is that CBS 19 is a young, pre-main sequence (PMS)
star surrounded by a warm, circumstellar disk that causes excess nIR
emission. In fact, after accounting for the extinction derived from
$N_{\rm H,X}$, the location of CBS 19 in the nIR color-color diagram
(Fig.~\ref{fig_ccd}b) agrees with the locus of the classical T\,Tauri
stars (CTTS), a sub-class of PMS stars that are still actively
accreting from their circumstellar disk \citep{meyeea97}. The
prominent but relatively weak (compared to CVs) H$\alpha$ emission is
also in line with this classification.  The X-ray properties of CBS 19
are typical for PMS stars, where X-rays are generated by a high level
of magnetic activity, with a possible contribution of (relatively
soft) X-rays from the accretion process \citep[e.g.][]{feigea07}. We
note that the disk can be a source of local extinction, and therefore
our estimates of the distance ($d=3.0\pm0.8$ kpc) and luminosity
($\log L_{\rm X} = 31.8\pm0.3$ erg s$^{-1}$) based on the Drimmel
model may be too high.

The Li~I 6708\AA~line is a well-known indicator of youth. Detection of
this line in high-resolution optical spectra of CBS 19 would confirm
its T~Tauri nature.

\subsubsection{CBS 8: an RS\,CVn binary} \label{results_ab}

Our Hydra spectrum of the candidate counterpart to CBS 8 classifies it
as an early/mid K star.  The X-ray spectrum of CBS 8 is rather hard
for steady, stellar coronal emission. For a power-law model we find
$\Gamma=1.9 \pm 0.2$; for a bremsstrahlung model we find $kT =
5.5^{+2.3}_{-1.3}$ keV. The count rate is stable throughout the 48-ks
observation, so it is unlikely that the hard spectrum is the result of
a typical coronal flare which can have a decay time of up to $\sim$15
ks \citep{gued04}. CBS 8 appears in our database four more times,
always with a lower count rate (at a level of $>$3 $\sigma$). The
largest difference is found with respect to ObsID 3807 (2002 Sep 24,
23 ks GTI) where the source is detected with 2.7$\pm$0.4 cts ks$^{-1}$
compared to 10.7$\pm$0.5 cts ks$^{-1}$ in the observation analyzed
here, taken a year later. In ObsID 3807 the source appears softer,
with $E_{50}=1.2\pm0.1$ keV compared to $E_{50}=1.64\pm0.04$ keV. The
hardening and brightening of the source could point at an increased
level of coronal activity occurring on a $\sim$year time scale.

The distance derived from $N_{\rm H,X}$ gives an absolute magnitude
$M_V=0.4^{+1.2}_{-0.9}$, pointing to the star being a luminosity-class
III giant (for a K3\,III star, $M_V=+0.8$). Also from
Fig.~\ref{fig_ccd} it can be seen that the spectral type better agrees
with the optical/nIR colors if we assume the star is a giant instead
of a dwarf. The H$\alpha$ absorption line is very weak, a sign that
the line may be filled in by excess H$\alpha$ emission. We tentatively
classify CBS 8 as an active RS\,CVn-type binary.

\subsubsection{CBS 14: an ultracool dwarf?}\label{results_ultracool}

The X-ray spectrum of CBS 14 is well fitted with a 2$T$ MeKaL model
with $kT_1\approx0.9$ keV and $kT_2\approx2.5$ keV, and minimal
absorption ($N_{\rm H,X} < 2\times10^{20}$ cm$^{-2}$). Apart from
small-amplitude variations, the count rate steadily declines during
the $\sim$9.5 ks observation. The spectrum becomes softer when the
source gets fainter (Fig.~\ref{xlc_cbs4_14}). This is typical for a
coronal flare, but since the source was already bright at the start of
the observation it is impossible to say if the light curve shows the
characteristic ``fast rise, slow decay''-profile of a flare.

\begin{figure}
\centering
\includegraphics[width=8.5cm]{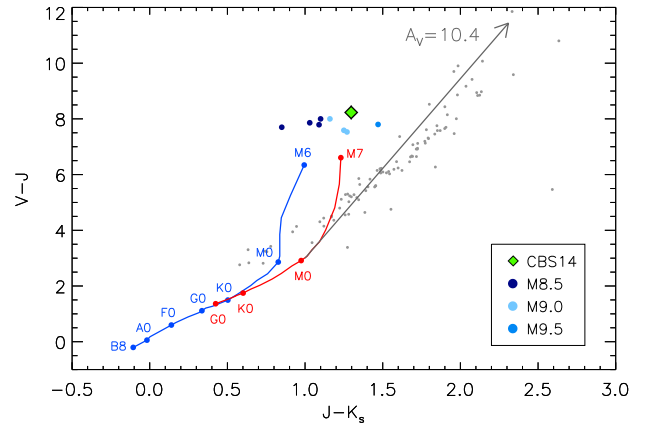}
\caption{Color-color diagram illustrating the tentative classification
  of CBS 14 as a very late-type dwarf based on the optical/nIR colors
  of its candidate counterpart ({\em green diamond}). For comparison
  we include the intrinsic colors of main-sequence stars as a black
  ({\em blue}) curve, of giants as a light gray ({\em red}) curve, and
  nearby ($\lesssim$30 pc) ultracool dwarfs ({\em filled circles};
  \cite{henrea04}).  Most stars in a 5\arcmin\,$\times$
  5\arcmin\,region centered on CBS 14 ({\em grey dots}) lie along the
  reddening vector \citep{nishea08}, shown with a length representing
  the total Galactic extinction in this direction ($A_V=10.4$). The
  candidate counterpart to CBS 14 lies away from this vector among the
  nearby ultracool dwarfs. Classification as an ultracool dwarf is
  consistent with the low extinction ($N_{\rm H,X}\lesssim 2~10^{20}$
  cm$^{-2}$ or $A_V\lesssim 0.1$) found from the X-ray
  spectrum. Errors on the colors of CBS 14 are smaller than the symbol
  size. {\em See the electronic edition of the Journal for a color
    version of this figure.}}
\label{fig_optnir_ccd}
\end{figure}

The error circle of CBS 14 contains an extremely red object with
$V-K_s=9.4$. Its optical/nIR colors place this object among the dwarfs
with spectral type M\,7 or later; these are the so-called {\em
  ultracool} dwarfs. This is illustrated in the color-color diagrams
of Figs.~\ref{fig_ccd} and \ref{fig_optnir_ccd}, which also include
nearby ultracool dwarfs. By comparing its $J$ magnitude, and $I-J$ and
$J-K_s$ colors with those of ultracool dwarfs at known distances
\citep{phanea08}, we estimate an approximate spectral type of M\,9 and
$d\approx24$ pc.

The area around CBS 14 is included in the field of view of the
EPIC-MOS cameras in two {\em XMM-Newton} observations: 0093670501
(2001 Mar 2; 14 ks) and 0207300201 (2004 Feb 22; 34 ks); the source
lies outside the field of the EPIC-PN camera. After filtering out
background flares, $\sim$13 ks and $\sim$15 ks of exposure time
remain. No source is seen at the location of CBS 14. Upper limits on
the flux are estimated by first extracting all counts (0.3--8 keV)
from within 20\arcsec~of the {\em Chandra} source position. Using
\cite{gehr86}, we compute the 3-$\sigma$ upper limit on the number of
counts. The background contribution is estimated from an annulus
around the source position. The resulting limit on the net source
counts is converted to a count rate limit using the value of the
exposure map at the source position. With the spectral model of
Table~\ref{spectral-param}, we estimate that the most restrictive
3-$\sigma$ flux limit comes from the 2001 observation and is $\sim$1.6
$\times$ 10$^{-14}$ erg s$^{-1}$ cm$^{-2}$. It is possible that the
spectrum is softer when the source is not flaring, but for a $kT=0.5$
keV MeKaL spectrum the upper limit on the flux changes only 10\%. The
flux from the {\em Chandra} observation is uncertain due to the
pileup. A conservative lower limit is obtained by assuming that the
detected count rate is the incident count rate. This implies an
unabsorbed flux of $>1.6\times10^{-12}$ erg s$^{-1}$ cm$^{-2}$ (0.3--8
keV), meaning that CBS 14 was $\gtrsim$100 brighter during the {\em
  Chandra} observation compared to the time of the {\em XMM-Newton}
observations. Therefore, our estimate of $(F_{\rm X}/F_{\rm O})_{\rm
  u}$ is very uncertain as our X-ray and optical observations were not
simultaneous.

CBS 14 was not detected in the {\em ROSAT} All Sky Survey. The
detection limit of $\sim$0.015 cts s$^{-1}$ \citep[0.1--2.4 keV,
][]{huenea98b} gives an upper limit to the intrinsic flux of $\sim$1.7
$\times$ 10$^{-13}$ erg s$^{-1}$ cm$^{-2}$ for a $kT=0.5$ keV MeKaL
model and $N_{\rm H,X}=2 \times 10^{20}$ cm$^{-2}$.  The lower limit
to the {\em Chandra} flux is $1.4 \times 10^{-12}$ erg s$^{-1}$
cm$^{-2}$ when extrapolated to the {\em ROSAT} band, implying a flux
increase of at least a factor of $\sim$8.

Spectroscopic follow-up is needed to test our tentative classification
of CBS 14 as an ultracool dwarf. It is not obvious what an alternative
explanation could be. The X-ray spectrum points at negligible
extinction and excludes the option of a background AGN or a
heavily-obscured object to explain the red optical/nIR colors. The
number of random interlopers inside the area searched for counterparts
is $N_{\rm ran}=0.5$, so there is a reasonable chance that the red
object is a chance coincidence. This implies that the true counterpart
has $R>24$.

\subsection{AGN}\label{results_agn}

Based on its optical spectrum, we classify CBS 2 as an AGN. Broad
emission lines, coming from gas moving at high speeds around the
central supermassive black hole, are prominently visible and constrain
the redshift to $z\approx1.78$.  Our database includes two detections
of CBS 2: the one analyzed here and a detection in ObsID 835 (2000 Jan
5; 26.5 ks GTI), in which the source is about half as bright
(4.9$\pm$0.5 versus 10$\pm$0.7 cts ks$^{-1}$). The energy quantiles
are consistent.

\subsection{AGNs or Galactic accreting binaries}\label{results_unknown}

The relatively hard X-ray spectra of CBS 1, 9, 13, 15, 16, and 18
suggest that their X-rays are accretion-powered, but with the
information we have we cannot distinguish between AGNs or Galactic
sources. The spectra of most AGNs can be described by power laws with
photon spectral index $\Gamma \approx 1-2.5$ \citep[see
  e.g.][]{tozzea06}. The spectra of our sources have photon indices
that lie in this range, except for CBS 18 (see below).

The large errors on $N_{\rm H,X}$ make it difficult to constrain the
distances, as the $N_{\rm H,X}$ values are consistent with the total
Galactic column densities $N_{\rm H,Gal}$ towards these sources at the
$<$2.5$\sigma$ level. Therefore, for these sources we can only derive
lower limits to the distance. Only for CBS 13 does $N_{\rm H,X}$
exceed $N_{\rm H,Gal}$ by $>$3$\sigma$; but while an AGN nature is the
most obvious explanation, one cannot exclude it is a highly-obscured
Galactic source.

For CBS 1 and 9 we have candidate optical counterparts, but no optical
spectra to classify them. The H$\alpha-R$ colors show no indication of
an excess H$\alpha$ flux.  Whereas the spectra of most CVs and
quiescent X-ray binaries show clear H$\alpha$ emission
(e.g.~\citealt{torrea04}), some qLMXBs have H$\alpha$ emission lines
with equivalent widths $> -10$\AA~(e.g.~\cite{elebea09}). Such weak
lines are not expected to make the H$\alpha-R$ color stand out (see
Fig.~4 in \cite{zhaogrindea05}).  We also note that our method to look
for H$\alpha$ excess sources only works for redshift $z=0$, and fails
for objects at significantly higher redshifts. Follow-up spectroscopy
of the candidate counterparts is necessary for an unambiguous
classification.  CBS 13, 15, 16, and 18 have no candidate optical
counterparts down to the limiting magnitude of our Mosaic images. The
(limits on the) $\log (F_{\rm X}/F_R)_{\rm u}$ values of all six
sources are consistent with both a Galactic and an extra-galactic
interpretation \citep[e.g.][]{hornea2001}.

For CBS 18 $N_{\rm H,X}$ is relatively low compared to $N_{\rm H,Gal}$
(($2.1\pm1.5)\times10^{22}$ cm$^{-2}$ versus $5.8\times10^{22}$
cm$^{-2}$), and it may be the best candidate among this subset to be
an accreting binary. The spectral fit gives an unusually flat photon
index ($\Gamma=-0.3\pm0.4$), and there is a hint of a systematic trend
in the residuals above 5 keV. The data are of insufficient quality to
test for the presence of an emission line, which is not included in
our model but could skew the spectrum to seem flatter than it actually
is.

CBS 18 is also detected in the consecutive {\em Chandra} observations
949 (42 ks) and 1523 (58 ks) from 2000 Feb 24/25. The energy quantiles
and count rates are consistent in all observations. The combined light
curve from ObsIDs 949 and 1523 shows a weak sign of periodicity at
$P=503\pm1$ s, with the corresponding peak in the Lomb-Scargle
periodogram just above the 99\% confidence level. As the folded light
curve does not look convincingly variable, we consider this detection
marginal at best. There is no sign of periodicity in the light curve
from the observation analyzed here.

CBS 18 lies in the field observed by \cite{ebisea05} and is their
source \# 200 or CXOGPE\,J184355.1-035829. Ebisawa et al.~did not find
a counterpart in the nIR follow-up campaign, which is complete down to
$J=18$, $H=17$, and $K_s=16$. The lack of a counterpart implies $\log
(F_{\rm X}/F_{K_s})_{\rm u} > 0.17$, which also points at an
accretion-powered source.

\section{Discussion}\label{discuss}

Our sample of 21 bright sources consists of 12 stars and 9
accretion-powered sources. Except for CBS 5, the stars are coronal
emitters where magnetic fields play a major role in developing or
sustaining hot plasmas. Among the accreting sources, two are Galactic
binaries (CBS 7 is a CV; CBS 17 is a candidate symbiotic binary), CBS
18 could be a CV, one is a confirmed AGN, and for the remaining six
this distinction is less clear. We first discuss three of our bright
sources in a broader context, and then continue to consider our sample
as a whole.

\subsection{Individual systems}\label{disc_sample}

\subsubsection{CBS 7: a likely magnetic cataclysmic variable}\label{disc_cbs7}

The hard spectrum of CBS 7 suggests it is a magnetic CV
\citep[e.g.][]{heinea08}, although some dwarf novae in quiescence have
hard spectra as well \citep[e.g.][]{balmea11}. Magnetic CVs can be
divided in two subclasses. In polars, the magnetic field is strong
enough ($B\gtrsim 10$ MG) to lock the spin to the orbital
motion. Intermediate polars (IPs) have fields of moderate strength
($B\approx 1-10$ MG), which are too weak to force synchronism. It is
not clear which of these classes CBS 7 belongs to. Our follow-up
optical and nIR data indicate that the longer of the two X-ray periods
is likely the orbital period ($P_b$) and we refer to
\cite{serveaastroph11} for a detailed discussion. The origin of the
shorter period is less clear. If the system is synchronized, the
variability could be caused by the two poles or accretion spots on the
white dwarf being partially (self)eclipsed. If the system is
asynchronous, the shorter period could reflect the white-dwarf spin
period ($P_s$), which would then be half the orbital period by
chance. The ratio of the periods ($P_s \approx 0.5 P_b$) is relatively
high compared to the typical ratio for IPs; most have $P_s \lesssim
0.3 P_b$. It would add CBS 7 to a small but growing number of
``near-synchronous'' IPs that may be evolving towards synchronism
\citep{nortea08,hongeaastroph11}. As the X-ray spectral properties of
CBS 7 do not vary with count rate, the dips are likely not caused by a
variation of the local amount of absorbing material. It is not
uncommon for magnetic CVs to be internally absorbed, which can cause
our distance and $L_{\rm X}$ estimates to be overestimated.

\subsubsection{CBS 17: a new hard X-ray emitting symbiotic? } \label{disc_cbs17}

Traditionally, symbiotic binaries were thought of as soft X-ray
emitters. This picture was mostly based on {\em ROSAT} observations
that were limited by its soft response \citep{muerea97}. Recently, an
increasing number of symbiotics has been detected at harder energies
\citep{kennea09,lunaea10,vdbergea06}. \cite{kennea09} suggest that for
RT\,Cru, T\,CrB, CD$-$57 3057, and CH\,Cyg the hard X-rays are thermal
and come from an accretion-disk boundary layer around a massive,
non-magnetic white dwarf. The potentially high white-dwarf masses
makes them interesting as candidate SN~Ia progenitors.  CBS\,17 could
be a member of this class of ``hard'' symbiotics.  Besides the hard
spectrum, similarities with this class include the weak optical
emission lines and possibly an internal absorption component. On the
other hand, our analysis suggests that the long-term variability is
not caused by variations in the column density, whereas in the systems
discussed by \cite{kennea09} the variations were found to result from
variations in the intrinsic absorption.

\subsubsection{The candidate ultra-cool dwarf CBS 14} \label{disc_cbs14}

Unlike earlier-type M stars that have a radiative core and convective
envelope, ultracool dwarfs are fully convective with cool and
effectively neutral atmospheres. Multi-wavelength diagnostics indeed
mark a change in magnetic activity around spectral type M\,7, but a
larger sample is needed to get a clearer picture: only 12 ultracool
dwarfs have been detected in X-rays \citep{bergea10,robrea10}. In this
context, it is important to establish the nature of CBS 14. Its colors
place it between the M\,9 and L\,0 dwarfs. So far, there is no firm
X-ray detection of a dwarf later than M\,9; the L\,2-dwarf Kelu-1 with
a marginal detection of 4 photons is the only possible exception
\citep{audaea07}. With $\log L_{{\rm 0.3-8 keV}}/L_{\rm bol} \gtrsim
-0.7$ (averaged over the observation, for a spectral type M\,9) and a
duration of at least 8 ks, CBS 14 could have been caught during one of
the strongest flares seen in ultracool dwarfs. The upper limit on the
quiesent luminosity from the {\em XMM-Newton} data is $1 \times
10^{27}$ erg s$^{-1}$ (0.3--8 keV). This is consistent with the limits
for other M\,9 stars but not very constraining as typical quiescent
luminosities are a few times 10$^{26}$ erg s$^{-1}$
\citep{bergea10,robrea10}

\subsection{Overall sample} \label{disc_logNlogS}

\subsubsection{Contribution from background sources} \label{disc_agn}

We estimate the expected number of AGNs in our sample using the
cumulative X-ray point-source density as a function of flux ($\log N -
\log S$ distribution) derived for high Galactic latitudes.  To this
end, we determine the flux limit of an observation by calculating the
flux of a source observed at the aimpoint with 250 net counts (0.3--8
keV). Here we assume a $\Gamma=1.7$ power-law spectrum and a column
density equal to the integrated Galactic $N_{\rm H}$ along the line of
sight. We do this for each of the 63 observations that we searched for
bright sources; the values for the observations analyzed here are
included in column 10 of Table~\ref{x-sample-table}. The average flux
limit of an observation is higher than the value thus computed as the
sensitivity decreases with offset angle from the aimpoint. As a
first-order correction for the vignetting, we multiply the flux limits
for the aimpoint by the ratio of the value of the exposure map at the
aimpoint and the mode of the exposure map. This gives a correction
factor of 1.09 for ACIS-S, and of 1.04\footnote{ Fig.~6.6 in the {\em
    Chandra} POG suggests that this factor can be $\sim$25\% higher
  for far off-axis ($\gtrsim$10\arcmin) ACIS-I detections, depending
  on the spectrum of a source.} for ACIS-I. We find that the number of
AGNs predicted is rather uncertain. The $\log N - \log S$ curves
(0.3--8 keV) from \cite{kimea07} predict a total of 19--57 AGNs, where
the range is defined by the $2\sigma$ error margins in the parameters
of the $\log N - \log S$ equation. If we use the extinction maps from
\cite{schlea98}, which on average predict a higher integrated Galactic
column density, the number of expected AGNs is 13--42. In fact, we
have 1 confirmed AGN and 6 candidate AGNs (\S\ref{results_agn},
\S\ref{results_unknown}). However, if we take into account statistical
errors (which are not included in the ranges given above), and the
possibility that the Galactic extinction is still underestimated
(neither the Drimmel nor the Schlegel extinction models has been
verified externally in an extensive way), the predicted and observed
number of AGNs are not so discrepant as they might seem at first
sight. In any case, it is likely that most of the sources in
\S\ref{results_unknown} are indeed AGNs.

\subsubsection{Source density and $\log N - \log S$ curves}

\begin{figure}
\centering
\includegraphics[width=7.5cm]{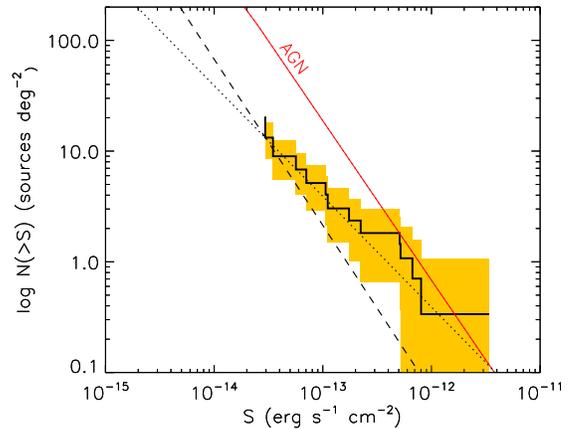}
\caption{Cumulative source density versus flux (0.3--8 keV) for {\em
    only the Galactic sources} in our sample. The relations $ \log
  N(>S) \propto S^{\alpha}$ with $\alpha=-1$ ({\em dotted line}) and
  $\alpha=-1.5$ ({\em dashed line}, expected slope for an isotropic
  distribution), each normalized to the faintest flux in our sample,
  are also included. For comparison, we also show the $\log N - \log
  S$ curve for AGNs from Kim et al.~(2007; {\em solid red line}),
  which indeed has a slope close to $\alpha=-1.5$. Correction for
  Galactic absorption was done with the model from \cite{drimea03};
  the model from \cite{schlea98} gives a consistent result. {\em See
    the electronic edition of the Journal for a color version of this
    figure.}}
\label{fig_lognlogs}
\end{figure}

Using our classifications to separate the Galactic sources in our
sample from the extra-galactic sources (including the six AGN
candidates), we have computed the cumulative source density versus
flux for the Galactic sources only. We do this by adding up the
contribution of each source and normalizing it by the area in which it
could have been detected based on the flux limits of each of the 63
observations searched. As in this case we only consider Galactic
sources, we have used a different spectral model to calculate the flux
limits than in \S\ref{disc_agn}. To account for the typically softer
spectra of the Galactic sources (see~Table~\ref{spectral-param}), we
use a power law with photon index $\Gamma=3$ and correct for only 25\%
of the integrated $N_{\rm H}$ along the line of sight. This value of
$\Gamma$ is only appropriate for coronal sources, but we find that
$\Gamma=1.7$ and correcting for the total integrated $N_{\rm H}$ give
the same results. The $\log N - \log S$ curve for the 0.3--8 keV band
is shown in Fig.~\ref{fig_lognlogs}. At the high end of the flux range
($\gtrsim 5 \times 10^{-13}$ erg s$^{-1}$ cm$^{-2}$) lie the accreting
sources CBS 7 and CBS 17, the early-type star CBS 5, the highly
variable source CBS 14, and the stellar coronal source CBS 10. The
nine sources below this limit are all likely late-type active stars.

We do not have enough statistics to do a detailed analysis, but can
make a qualitative statement. The slope $\alpha$ of the curve $\log N
(>S) \propto S^{-\alpha}$ appears flatter than the slope for an
isotropic distribution ($\alpha=-1.5$) that is expected for a truly
homogenous angular distribution like that of AGNs, or for a population
of local sources that is sufficiently close as to seem isotropic. The
slope is closer to the value expected for a disk distribution
($\alpha=-1$). The derived distances indeed place our Galactic sources
up to 3--5 kpc away (Table~\ref{flux-dist-lum}).  Other Galactic plane
surveys find a similar flattening of the Galactic $\log N - \log S$
distributions \citep[e.g.][]{hertgrin84}, which for the XGPS is mostly
accounted for by soft stellar coronal sources \citep{motcea10}. The
resulting Galactic source density is roughly similar to the value from
the XPGS: we find about 6--15 sources deg$^{-2}$ above a limiting flux
of $5\times10^{-14}$ erg s$^{-1}$ cm$^{-2}$ (0.3--8 keV) compared to
$\sim$3 sources deg$^{-2}$ (0.4--2 keV) and $\sim$15 sources
deg$^{-2}$ (2--10 keV) in the region studied by the XGPS
\citep{motcea10}. Galactic sources are outnumberd by AGNs above this
flux limit (by a factor of 3--7 according to \cite{kimea07}).

\subsubsection{Source templates}

In this study we have avoided the central, most obscured part of the
plane, which enabled us to use the properties of the optical and nIR
counterparts for source classification. Would {\em Chandra} have
detected our sources if they were located in the central bulge, and if
so, can we use them as templates to constrain the nature of more
distant obscured sources?  We answer this question for the specific
case of the 10\arcmin~$\times$ 10\arcmin~region around Sgr A*---the
Galactic center region or GCR---where {\em Chandra} found a large
number (thousands) of X-ray point sources, most of which remain
unidentified at other wavelengths. We apply the canonical extinction
for the GCR of $N_{\rm H,X}=6 \times 10^{22}$ cm$^{-2}$ to our
intrinsic source fluxes, and put them at a distance of 8.5 kpc. The
deep pointings of the Sgr A* field reach a sensitivity of at least
$5\times10^{-15}$ erg s$^{-1}$ cm$^{-2}$ ($S/N \geq 3$ between 2--8
keV for 700 ks of stacked exposures; \cite{hongvandea09}). All
(candidate) AGNs except CBS 2, the two (likely) Galactic
accretion-powered sources CBS 7 and CBS 17, and the RS\,CVn binary CBS
8 would fall above this X-ray limit. Only two sources would be bright
enough in the nIR to be detectable in our GCR $K_s$ band observations
\citep{laycea05}, viz.~CBS 8 and CBS 17. With $K_s\approx15$ and 14.1,
respectively, they would lie above the confusion limit of $K_s=15.4$
in most of the region, and $K_s=14.5$ in the most crowded region
within 1\arcmin~of Sgr A*. With higher-resolution images, obtainable
with AO imaging, they would be easily detectable. However, as their
spectra do not show strong signatures of activity or accretion (not in
the optical, at least), it is difficult to distinguish them from the
random coincidences with late-type evolved stars, which abound in the
old bulge. While \cite{laycea05} showed that at most 10\% of the
unidentified sources around Sgr A* have such bright counterparts, it is
important to recognize that it is not only young wind-fed quiescent
Be-HMXBs, which have received much of the focus so far, that
contribute to this nIR ``bright'' population of X-ray sources, but
also active binaries of the RS\,CVn-type and symbiotic binaries.

\section{Future work}

The present work is limited to a small sample of bright sources, but
has uncovered a number of interesting systems that are worth detailed
follow-up. Some of it is already underway. \cite{serveaastroph11}
present the results of an optical/nIR study of CBS 7, and optical
spectroscopy to establish the nature of CBS 14 is planned. An
extension of this work to the entire ChaMPlane database, which is
almost 3 orders of magnitude larger, is guaranteed to find more
individual sources of interest. It also allows us to construct deeper
$\log N - \log S$ curves in distinct sections of the plane with enough
statistics to investigate differences in density and distribution of
various types (accreting versus non-accreting) of faint Galactic X-ray
sources, and can contribute to our understanding of the composition of
the resolved part of the Galactic Ridge. Whereas the present study
falls short in the number of sources included, source classification
of the larger sample has to be done in a more statistical sense than
we were able to do for the bright sources.

\acknowledgements 
This work was supported by NSF grants AST-0098683, {\em Chandra}
grants GO3-4033A and GO6-7088X, and includes work carried out (by
K.~Penner) as part of the Research Experience for Undergraduates
program at the Harvard-Smithsonian Center for Astrophysics. We thank
J.~Hernandez for providing the {\tt SPTCLASS} code, and K.~Stassun for
discussing the optical and near-infrared properties of young stars. We
also thank the {\em Chandra} X-ray Center for support.

{\it Facilities:} \facility{CXO}, \facility{XMM},
\facility{Magellan:Baade (IMACS)}, \facility{FLWO:1.5m (FAST)},
\facility{Mayall (Mosaic)}, \facility{Blanco (Mosaic, Hydra)},
\facility{WIYN (Hydra)}


\end{document}